\documentclass[usenatbib]{mn2e}
\usepackage{times}
\usepackage{epsfig}
\usepackage{amsmath}
\title[]{The colour of galaxies in distant groups}
\author[Balogh et al.]{Michael L. Balogh$^{1}$, Sean L. McGee$^{1}$, Dave
  Wilman$^{2}$, Richard G. Bower$^{3}$,
\newauthor George Hau$^{4}$, Simon  L. Morris$^{3}$,
J.~S. Mulchaey$^{5}$,  A. Oemler Jr.$^{5}$, Laura Parker$^{6}$,
\newauthor Stephen Gwyn$^{7}$
\\
$^{1}$Department of Physics and Astronomy, University of Waterloo, Waterloo, Ontario, N2L 3G1, Canada\\
$^{2}$Max--Planck--Institut f{\" u}r extraterrestrische Physik, Giessenbachstrasse 85748 Garching Germany\\
$^{3}$Department of Physics, University of Durham, Durham, UK, DH1 3LE\\
$^{4}$Centre for Astrophysics and Supercomputing, Swinburne University of Technology, Hawthorn, Victoria 3122, Australia\\
$^{5}$Observatories of the Carnegie Institution, 813 Santa Barbara
Street, Pasadena, California, USA\\
$^{6}$Department of Physics and Astronomy, McMaster University, Hamilton, Ontario, L8S 4M1 Canada\\
$^{7}$Canadian Astronomical Data Centre, Herzberg Institute of Astrophysics Victoria, BC, Canada\\
}
\date{\today}

\def\gtrsim{\mathrel{\raise0.35ex\hbox{$\scriptstyle >$}\kern-0.6em
\lower0.40ex\hbox{{$\scriptstyle \sim$}}}}
\def\lesssim{\mathrel{\raise0.35ex\hbox{$\scriptstyle <$}\kern-0.6em
\lower0.40ex\hbox{{$\scriptstyle \sim$}}}}

\def\k04{{^{0.4}K_s}}
\def\s04{{^{0.4}S1}}

\begin{document} 
\maketitle
\begin{abstract}
We present new optical and near-infrared imaging for a sample of 98
spectroscopically-selected galaxy groups at $0.25<z<0.55$, most of which have velocity dispersions
$\sigma<500$~km/s.  We use {\it psf}-matched aperture photometry to measure
accurate colours for group members and the surrounding field
population.  The sample is statistically complete above a stellar mass
limit of approximately $M=1\times 10^{10}M_\odot$.  The overall colour
distribution is bimodal in both the field and group samples; but at
fixed luminosity the fraction of group galaxies populating the red peak
is larger, by $\sim 20\pm7$ per
cent, than that of the field.  In particular, group members with early-type morphologies, as identified
in {\it Hubble Space Telescope} imaging, exhibit a tight red sequence,
similar to that seen for more massive clusters.  Using optical and near-infrared
colours, including data from the {\it Spitzer Space Telescope}, we show that approximately 20--30 per cent of galaxies on the
red sequence may be dust-reddened galaxies with non-negligible star
formation and early-spiral morphologies.  This is true of both the
field and group sample, and shows little dependence on near infrared
luminosity.  Thus, the fraction of bright ($^{0.4}M_{K}<-22$) group members with no sign of
star formation or AGN activity, as identified by their colours or [OII]
emission, is $54\pm 6$ per cent.  Our field sample, which
includes galaxies in all environments, contains $35\pm3$ per cent
of such inactive galaxies, consistent with the amount expected if {\it all}
such galaxies are located in groups and clusters.
This reinforces our earlier conclusions,
that dense environments at $z\lesssim 0.5$ are associated with a premature cessation
of star formation in some galaxies; in particular we find no evidence for significantly enhanced star
formation in these environments.  Simple galaxy formation models
predict a quenching of star formation in groups that is too efficient,
overpopulating the red sequence.  Attempts to fix this by increasing
the timescale of this quenching equally for all group members distorts the colour distribution in a
way that is inconsistent with observations.  
\end{abstract}

\section{Introduction}\label{sec-intro}
The evolution of the galaxy population since a redshift of about $z=1$
appears to be a strong function of both stellar mass and environment
\citep[e.g.][]{Juneau+04,BB04,Blanton04,BaldryV,Haines+06,HGM08,vdB07_short,GB08,CB08}.
The overall effect is for star formation activity to decrease with
time, with such activity apparently ceasing first in the most massive
galaxies.  Fundamentally, this evolution is linked to the accretion,
radiative cooling, and reheating of gas by individual galaxies
\citep{Bower05,Croton05,DB06,Somerville08}.  But a wide
variety of physical processes operating on an enormous range of scales
may play important roles in how this gas component behaves, and we are
still a long way from knowing which are
most important on which scales, and at which epochs.

A particularly efficient way to track the evolution of star formation
activity is through two- or three-colour optical photometry.  Optical
colours are almost equally sensitive to galaxy age, metal abundance,
and dust content, which at first seems discouraging.  To first order,
however, the bulk of the galaxy population appears to be driven by only
a few, correlated parameters.  Most markedly, galaxies can be cleanly
divided into two families --- one red and one blue --- that differ mostly in their
star formation history \citep[e.g.][]{Strateva01_short,Baldry03}; this
distinction appears to hold to at least $z=2$
\citep{BellGems2,Gerke06,Cassata08}.  Within each of these families,
there is a well-known correlation with magnitude (or stellar mass),
that is at least partly driven by a systematic variation in metal
content and possibly dust.  The origin of the significant scatter about
these correlations is not well understood, but almost certainly
includes variation in all three parameters.  In particular, it is known
that the red population includes galaxies with significant amounts of
star formation \citep[e.g.][]{cnoc2_irac,STAGES-dust}.

Making a few important assumptions, the stellar mass of a galaxy can be
relatively easily measured from its luminosity.  Using large-area redshift
surveys and multicolour imaging (including the near-infrared) it has
therefore become possible to study correlations of galaxy properties with mass, over a long redshift
baseline \citep[e.g.][]{CB08}.  Although still a subject of debate, the
role of mass or surface mass density as a driving factor of galaxy
evolution is becoming clear \citep{Taylor+08,M+08}.   The stellar mass
functions of red and blue galaxies are distinct, and they
also appear to evolve differently with redshift
\citep[e.g.][]{Bell+07_short}.  In particular, the
most massive galaxies ($M\gtrsim 2\times 
10^{10}M_\odot$) today have predominantly old stellar populations,
independent of environment \citep[e.g.][]{Kauffmann-SDSS1_short}.  
It has been proposed that the supermassive black
holes harboured within these galaxies can provide enough feedback energy to reheat or expel the
cooling gas necessary for star formation to occur
\citep{BD03,Croton05,Bower05,DB06,Sijacki07}.   Although the giant
galaxies found in clusters can sometimes show evidence for recent star
formation, especially if they are located near the peak of an X-ray
cool core \citep[e.g.][]{C+99,Edwards+07,BHBM08,RMN08}, the associated
star formation rates are still low relative to the galaxy's mass \citep[e.g.][]{OD+08}.

On the other hand, the role of environment, if any, is still unclear.
It is well known that dense environments today 
harbour less active galaxies, in general \citep[e.g.][and many
others]{B+97,2dF_short,Sloan_sfr_short,Blanton04,Haines+06}.  That correlation is weakened
somewhat when one accounts for the fact that denser environments
contain more massive galaxies \citep[e.g.][]{BaldryV}, but it still
persists; in fact, the data show that environmental effects are most
easily observed in the low-mass galaxy population
  \citep[e.g.][]{BaldryV,vdB07_short}.  

In principle, the evolution of galaxies in dense environments can be
observed directly by comparing samples at different redshift.  This is
complicated however by the hierarchical growth of structure, which
means that galaxies within clusters today were found within systems
spanning a wide range of lower masses at higher redshift \citep[e.g.][]{Berr+08,McGee-accretion}. 
Locally, the
properties of galaxies at fixed stellar mass show little or no
dependence on the mass of the group or cluster in which they reside, at
least for systems with velocity dispersions
$\sigma>500$~km/s
\citep[e.g.][]{2dfsdss,Weinmann_1+06,Pogg05,Pogg+08,Finn08,P+08}.
These massive systems also appear to evolve little with redshift, at least for
$z\lesssim1$ \citep[e.g.][]{Nakata,P+08}.  On the other hand, lower-mass
groups may be generally more active at higher redshift \citep[e.g.][]{Pogg+08}, though
  always less so than the general field
  \citep[e.g.][]{CNOC2_groupsII,JMLF}. If these intermediate redshift
  groups at $z\sim 0.5$ are to evolve into more massive clusters by the
  present day, this implies that there
  is a mechanism for truncating star formation in systems with halo
  masses approximately corresponding to those of small galaxy groups \citep{GB08}.
At even higher redshifts, $z\sim 1$, 
  although dense environments still contain a higher frequency of
  evolved galaxies, those that {\it are} active in groups and clusters may have
  greater star formation rates than the field, implying that local
  interactions may be more important than the global environment
  \citep[e.g.][]{DEEPII_envt,DEEPII_envt_again,Elbaz+07,Muzzin+08}. 

The simplest model for environmental effects that has shown some
success is one where satellite galaxies (those not at the centre of the
potential) are immediately stripped of all hot gas
\citep[e.g.][]{WF91,Cole2000}.  This, coupled with fairly short gas
consumption and ejection timescales, leads to rapid truncation times, and satellite
populations that are uniformly red at the present day
\citep{Croton05,Bower05}.  However, some observations suggest that the
quenching is a more gradual process
\citep[e.g.][]{PSG,infall,Moran+06,Gallazzi+08,McGee,McGee-accretion}.  One clue to why
this might be is that some active galaxies
in dense environments exhibit truncated or distorted
  H$\alpha$ disks, implying that star formation is truncated first in
  the outskirts of the galaxies \citep[e.g.][]{KK04,BMA}, rather than
  all at once.  The result
  is that nearby observed groups and clusters tend to have many more blue,
  active low-mass galaxies than are predicted by this simple gas
  stripping model \citep{Weinmann+06_short,BaldryV,GB08,Font+08}.

The observations above suggest that galaxy groups at $z\approx 0.5$ may
represent a critical environment for galaxy evolution.  We have
therefore embarked on an ambitious programme to obtain a complete
census of the galaxy population in these environments.  Our survey is
based on the CNOC2 redshift survey \citep{CNOC2-I_short}, as described
in \citet{CNOC2_groupsI} and \citet{CNOC_groups2}.  In a series of
papers we have shown that the galaxies in these groups do not differ
strongly from the surrounding field, except that they have a slightly larger fraction of
passively--evolving, bulge-dominated galaxies
\citep{CNOC2_groupsII,cnoc2_ir,cnoc2_irac,McGee}. However, these works
were hampered by the relatively poor quality of the original optical
imaging of the CNOC2 survey.  Here we present improved optical imaging, in the
  $ugrizBVRI$ filters from the CFHT Megacam and CFH12K imagers.  These
  are combined with our ground-based and {\it Spitzer} IRAC
  near-infrared (NIR) photometry \citep{cnoc2_ir,cnoc2_irac}, including
  new $K_s$ observations from the New Technology Telescope (NTT).

We describe the survey data in \S~\ref{sec-data}, including a summary
of the original survey observations and our follow-up spectroscopy.
The analysis of the new imaging data is presented in
\S~\ref{sec-anal}.  The results are described in \S~\ref{sec-results},
and discussed in \S~\ref{sec-discuss}.  Our conclusions follow, in \S~\ref{sec-conc}.
For all cosmology--dependent calculations, we assume $\Omega_m=0.3$, $\Omega_\Lambda=0.7$
and $h=0.7$.  All magnitudes are in the AB system unless otherwise noted.

\section{The Data}\label{sec-data}
The original CNOC2 redshift survey consists of four widely separated
fields, at approximate RA positions of 2h, 9h, 14h and 21h.  Each patch was
surveyed with a mosaic of several $9.3\arcmin\times8.3\arcmin$ CFHT-MOS fields, in an unusual and
somewhat awkward geometry, consisting of a main square ($3\times3$ MOS
fields) area, and two or three linear extensions
\citep{CNOC2-I_short}.  Medium-resolution optical spectroscopy was
obtained for $\sim 6000$ galaxies in these fields; target selection was
based only on $R_c$ magnitude, and the spectroscopic completeness is 50 per cent
at magnitudes brighter than $R_c=21.5$.  From these data,
\citet{CNOC_groups} identified several hundred groups over the redshift
range $0.1<z<0.55$, using a redshift-space friends-of-friends algorithm
with the additional criterion that the local overdensity be positive.  
We used the multi-object
spectrograph LDSS2 on the 6.5-m Baade telescope at Las Companas
Observatory in Chile to obtain a deeper and more complete sample of
galaxies in the region of twenty-six of these groups at $0.3\leq z\leq 0.55$
\citep{CNOC2_groupsI}.  We selected galaxies from the original MOS
imaging, and obtained $418$ additionalredshifts, of which 240 are in
the range $21\leq R_c \leq 22$ and 86 are confirmed members of identified
groups.  The result is a 74 per cent completeness for $R_c<22$, with
well-understood selection functions (that depend almost entirely on
$R_c$ magnitude and geometrical position) as described in
\citet{CNOC2_groupsI}.   The original CNOC2 spectra span a wavelength range
$4387$--$6285$\AA\ due to use of a band-limiting filter, while our
follow-up Magellan spectra cover the full optical range
$3700$--$8000$\AA. The redshifts have a typical uncertainty of
$100$km/s from the original survey, and $140$ km/s in our follow-up spectroscopy.

In this paper we present {\it psf}-matched aperture photometry at optical
and near-infrared wavelengths, for galaxies with redshifts in the four
CNOC2 survey fields.  Most of the
follow-up data we consider here is centred on the $25\arcmin\times25\arcmin$
contiguous central region,
and coverage of the ``extensions'' is more limited. The images considered here come from a mixture of
proprietary and archive data: {\it ugriz} and {\it BVRI} from the
Canada--France--Hawaii Telescope; $K_s$ from the William Herschel
Telescope and the New Technology Telescope; and 3.6$\mu m$
and 8$\mu m$
images from {\it Spitzer Space Telescope}.  These data and their
processing are described in more detail below.

\subsection{Optical imaging}\label{sec-optim}
The original redshift survey was based on relatively poor quality MOS
imaging.  Although perfectly acceptable for the spectroscopic selection
that was the focus of the original survey, statistical and systematic
uncertainties on the order of $\sim 0.1$ mag make it difficult to
cleanly identify distinct galaxy populations from their optical colours
\citep[for example, see Figure 10 in][]{cnoc2_ir}.  
Therefore, we have obtained additional optical imaging from two wide-field
mosaic cameras on the CFHT: the CFH12K and Megacam.  

We also use data from the Advanced Camera for Surveys (ACS) on the {\it
  Hubble Space Telescope}.  These data, described in \citet{McGee},
consist of 20 ACS fields in the F775W filter.  We will use the
morphological measurements from these data, as described in
\citet{Wilman_morph}, but not the photometry. 

\subsubsection{CFH12K images}
The CFH12K is the previous generation large imager on the CFHT, with 12
close-packed $2048\times 4096$ pixel CCDs, covering $42\times 28$ arcmin with 0.206\arcsec\
pixels.  We obtained most of these images from the CADC archive, and
they were subsequently 
processed and stacked as described below.  The exceptions
are the I-band images  in the 14h and 21h patches, which were obtained
and reduced as described in \citet{PHCH}.  Although some of the patches
have multiple pointings, covering a wider field, we only consider the single
CFH12K pointing near the centre of each
CNOC2 patch; dithering increases the coverage by a few arcmin in
each dimension.   All four patches have coverage in all four filters, $BVRI$.

\begin{table*}
\begin{tabular}{lllllll}
\hline
Patch&RA      & Dec  & Instrument & Filter & Total Exposure & Seeing\\
     &(J2000) & (J2000) &         &        & (ks) & (arcsec)\\
\hline
2h& 02:26:03.906& +00:19:34.69&CFH12K &B& 11.82 & 0.83\\
  &             &             &       &V& 2.65 & 0.75\\
  &             &             &       &R& 1.8 & 0.64\\
  &             &             &       &I& 1.8 & 0.55\\
  &             &             &Megacam&u& 13.97 & 0.80\\
  &             &             &       &g& 0.24 & 0.75\\
  &             &             &       &r& 0.48 & 0.66\\
  &             &             &       &i& 0.50 & 0.53\\
  &             &             &       &z& 0.36 & 0.58\\
\hline
9h& 09:23:46.657& +37:05:23.32&CFH12K &B& 12.96 & 0.82\\
  &             &             &       &V& 6.0 & 1.01\\
  &             &             &       &R& 19.08 & 0.82\\
  &             &             &       &I& 5.16 & 0.89\\
  &             &             &Megacam&g& 0.24 & 0.83\\
  &             &             &       &r& 0.48 & 0.78\\
  &             &             &       &i& 1.05 & 1.20\\
  &             &             &       &z& 0.36 & 0.66\\
\hline
14h&14:49:38.084& +09:08:57.62&CFH12K &B& 3.6 & 0.85\\
  &             &             &       &V& 3.6 & 0.94\\
  &             &             &       &R& 3.6 & 1.04\\
  &             &             &       &I& 14.4 & 0.75\\
  &             &             &Megacam&u& 16.64 & 0.80\\
  &             &             &       &g& 0.48 & 1.02\\
  &             &             &       &r& 0.96 & 0.91\\
  &             &             &       &i& 1.55 & 0.73\\
  &             &             &       &z& 0.72 & 0.72\\
\hline
21h&21:51:20.350& -05:33:26.77&CFH12K &B& 7.08 & 0.78\\
  &             &             &       &V& 7.44 & 0.76\\
  &             &             &       &R& 6.0 & 0.70\\
  &             &             &       &I& 14.4 & 0.68\\
  &             &             &Megacam&u& 14.72 & 0.80\\
  &             &             &       &g& 0.744 & 0.76\\
  &             &             &       &r& 0.528 & 0.69\\
  &             &             &       &z& 0.396 & 0.63\\
\hline
\end{tabular}
\caption{Optical observations obtained from the CFHT archive.
\label{tab-opt}}
\end{table*}
Individual exposure times are typically 540-840s, depending on the
filter; the total exposure times of the stacked images, and the
characteristic full-width at half maximum (fwhm) of the seeing, are given in
Table~\ref{tab-opt}.   Individual frames were dithered so the stacked frames suitably fill in
all chip gaps.  The {\sc Elixir}\footnote{http://www.cfht.hawaii.edu/Instruments/Elixir/home.html}
pipeline-processed data 
available from the CFHT archive were not directly
usable, as there are zeropoint shifts remaining from chip to chip that
are not correctly accounted for.  These shifts appear to be constant
for a given run, so it was possible to correct by comparing with
overlapping Sloan Digital Sky Survey (SDSS) data, where available.  However, there remains
small-scale background 
variation which affects photometry at the 0.03 mag level.  The only
fringe correction is done on the individual images by the {\sc Elixir}
pipeline.  

Those images overlapping with SDSS were calibrated by
converting SDSS magnitudes 
of stars to CFH12K magnitudes using the following empirical
conversions:  
\begin{align}
B =& g+0.4289(g-r)\\
V =& g - 0.5561(g-r) +0.0720(r-i)\\  
R =& r - 0.2178(r-i) - 0.0327(g-r)\\
I =& i -0.4451(i-z)\\
Z =& z - 0.1682(i-z).
\end{align}
These conversions have a scatter
of typically 0.01 mag (0.04 mag for $B$).  Stars in SDSS are thus
transformed on to the CFH12K system, and the images are calibrated
against this sample.  
As the 21h field does not overlap with the SDSS, we calibrated the
photometry by comparing with both \citet{Hsieh+05} and the original CNOC2
MOS photometry \citep{CNOC2-I_short}.  The zeropoints calibrated in this way are accurate to
approximately $0.03$ mags.

The 14h and 21h field I band images are considerably deeper, and were
processed and stacked as described in \citet{PHCH}.  

\subsubsection{Megacam images}
Public archival Megacam images in {\it griz} were found in the CADC archive, and
processed with {\sc Megapipe} \citep{Megapipe}.   Only the $i$ image for the 21h field was
not public at the time this work was done, and is therefore not included here.
These are shorter (5-10 min) exposures than the CFH12K data, but cover
a wider field and have a more accurate photometric and astrometric
solution.  However, as there are often only single pointings available,
there are significant chip gaps which limit our coverage.

The $u$ band data were obtained by us in semesters 2005A/B,
for all patches except the 9h field. 
There were two sets of six dithered images for each of the three
pointings; however, the dither pattern was not large enough to
completely fill all the chip gaps.  Each image was processed by the {\it Elixir}
v.2.0 pipeline, which detrends each of the 36 chips. This process includes
overscan, bias, masking and flat-fielding.  Only data taken under
photometric conditions were combined, using the {\sc swarp} software
provided by Terapix\footnote{http://terapix.iap.fr/}.  The combined, final
exposure times are listed in Table~\ref{tab-opt}.

\subsection{Near-infrared data}\label{sec-nir}
Near-infrared observations of these fields have been published by
us in \citet{cnoc2_ir}.  Here we summarize these data, and include
newer data obtained from the NTT.
For this paper, all of these 2MASS--calibrated magnitudes were
converted to the AB system by adding 1.85. 

\subsubsection{WHT INGRID observations}
Our first infrared data in these fields
were obtained at the {\it William Herschel Telescope} in March 2001, using the 
INGRID instrument.  Good data were obtained for 23
4.5\arcmin$\times$4.5\arcmin\ fields centred on
optically-detected galaxy groups in the 14h and 9h patches, with total
exposure times of about 12 minutes per field.   The data were reduced using the {\sc ipipe} NIR reduction pipeline
\citep{Gilbank03}, as described in  \citet{cnoc2_ir}\footnote{Small
  zeropoint adjustments have been made to three of the images since that
  work.}.  Our magnitude zeropoints (on the Vega system) and astrometric solution were calibrated using the
Two Micron All Sky Survey \citep[2MASS; see][]{2MASS}
point-source catalog $K_s$-band 4\arcsec\ standard aperture
magnitudes.  Due to the small field size, the photometric calibration
is generally based on only a few (1-3) bright stars, and this dominates
our uncertainties.
The seeing for all 23 fields is better than 1\arcsec.

\subsubsection{\it Spitzer IRAC}
Space-based NIR data from {\it Spitzer} IRAC was obtained from the archive,
for the 2h, 9h and 14h fields.  We used the processed 3.6$\mu m$ and
8$\mu m$ data
directly from the archive, with no further reduction.  These images
cover a relatively large area, of 36.6\arcmin$\times$43.6\arcmin.
However, they still do not generally cover the entire CNOC2 survey
area.  These data were also used and described in \citet{cnoc2_ir} and \citet{cnoc2_irac}.

\subsubsection{NTT SOFI observations}
New data were obtained for 27 6.9\arcmin$\times$6.9\arcmin\  fields in the 2h, 14 and 21h
patches, using the
SOFI instrument\footnote{Based on observing program 076.A.0346 and
  077.A.0224.} on the {\it NTT}, using both service mode (period 76)
and visitor mode (period 77).  Data were reduced with the NTT observatory pipeline.
An approximately 1 per cent gradient in the sky background following
dome flat fielding was corrected using an illumination correction.  The
image scale is 0.288 arcsec per pixel, and the gain/readnoise are 5.4
and 11.34$e^{-}$, respectively.  Fields were typically observed with 24--36
separate exposures which were then combined using the pipeline
software, for a total integration time of about 1 hour per field.  In some cases, better results were obtained by combining
subsets of the data first, and then coadding them after reduction.
The astrometric
solution was calculated based on the CFHT Megacam imaging where
available.  Outside the Megacam fields, either the ACS or original
CNOC2 astrometry was used.

The service mode run also includes short, photometric
exposures; however the visitor data was mostly obtained in
nonphotometric conditions.  In all cases, zeropoints were obtained by calibrating with the 2MASS point source
catalogue.  For the photometric observations taken in service mode, we
find this gives zeropoints for a given run that are consistent to
within 0.1 mag.  This is limited by the small number of stars in each field, and the
shallow depth of the 2MASS data. 
Comparing the SOFI and INGRID
data for the few cases where there is some overlap in sources, the {\it
  rms} of the magnitude difference is $\sim 0.2$ mag, implying the
INGRID zeropoint uncertainty is $\sim 0.17$ mag.

\section{Analysis}\label{sec-anal}
\subsection{Object Detection and Photometry}
Our primary goal is to obtain accurate colours for the galaxies in
the original CNOC2 survey with $R<22$, for which statistically complete
spectroscopy exists.  Our optical and near-infrared images are generally much deeper than
this.

Object detection was done on a combined image of R (CFH12K) and r
(Megacam).  We used {\sc swarp} to align and combine these images into
a single, large ``super'' image, using the {\sc LANCZOS3}
resampling option and an
Aitoff projection.  The advantages of using this image for object
detection are that the filter choice is close to that
used in the original CNOC2 survey selection, we achieve maximum depth
by combining the images, and we maximize continuity in the central
regions (since the Megacam data alone contains chip gaps).  The
disadvantage is that the detection limit is non-uniform; however, as
this is well below our limit of interest it is not a concern.  We note
that this approach also means there are survey regions with
spectroscopy and good data in other
filters (particularly the near-infrared) that are not included in our
catalogues because they do not have Megacam-$r$ or CFH12K-$R$ coverage.

All the other data, including the NIR and {\it Spitzer} images, were
aligned to this ``super'' image.  For the INGRID and SOFI data, we combine all
the individual, sky-subtracted images within a given patch, thus making a large image
of mostly empty pixels, with small regions of data.  

To measure accurate colours, we require all the images to be matched in
psf size.  To make optimal use of the data, this requires two
different data sets:

\noindent{\bf Optical/IR (OIR):} 
For the optical and ground-based NIR imaging alone, the
seeing is almost always better than 1\arcsec; the only major exception
is the 9h Megacam $i$ image, with 1.2 arcsec seeing.  Therefore, we
convolve all of these data with an appropriate Gaussian to achieve
1\arcsec\ resolution.  

\noindent{\bf OIR+IRAC:}  The IRAC data at 3.6 and 8 microns has a psf of approximately
2.3\arcsec, limited by the 1.2\arcsec pixel size.  Thus, whenever
  these data are used, colours must be measured in larger apertures and
  with a different psf correction.  To achieve this, we generate a
  second set of optical/IR images smoothed to 2.3\arcsec\ seeing.

Photometry was performed with {\sc SExtractor} v2.5 \citep{sextractor}, in two image
mode, where the ``super'' r-band image is always used for detection.
This ensures that the same centres are used for photometry in all
filters, and that both catalogues contain exactly the same objects, in
the same order.  RMS weight maps are used for both the detection and
measurement images.  Object detection is done using a 3\arcsec\
Gaussian kernel, and we require 3 pixels above the 1$\sigma$
threshold.  Deblending is performed with 32 sub-thresholds and a
minimum contrast of 0.002.  These parameters were chosen to provide the
best results above $R=22$, where we are most interested.  

Aperture photometry is performed using a local background annulus,
10\arcsec\ thick.  We use 3\arcsec\ diameter apertures for the OIR
catalogues (which are smoothed to 1\arcsec) and 5\arcsec\ apertures for
the OIR+IRAC catalogues (smoothed to 2.3 arcsec).  
In this paper, we always use the best (i.e. highest resolution) colour
when available.  
Colours are always computed using comparable magnitudes (i.e., measured
on images of the same resolution and using the same aperture sizes).
We have verified that there is
no systematic bias between colours measured in (e.g.) 3\arcsec\ and
5\arcsec\ apertures, only increased scatter in the latter.  Therefore, 
we will interchangeably use colours computed in different ways,
as necessary.  For example, $(r-i)$ colours will be based on the
5\arcsec\ aperture colours for the 9h field (driven by the poor seeing
of the $i$ images), but 3\arcsec\ apertures for the others.  There
  are obvious disadvantages to using fixed-aperture colours.  One is
  that colours are not measured within consistent fractions of the
  effective radius, so in the presence of gradients we could be
  introducing a systematic error with galaxy size.  However, we see no
  trend in the difference between 3\arcsec\ and 5\arcsec\ apertures with total
  luminosity, and conclude that this effect is small compared with our
  statistical uncertainties.  The other disadvantage is that colours
  will be incorrect for close pairs of galaxies, but these represent a
  relatively small fraction of our sample with $r<22$.  Thus, we
  conclude that fixed-aperture colours are well-suited to our purpose
  of identifying the main variations in galaxy population with
  environment and redshift.

For total magnitudes, we use the {\sc sextractor} MAG\_AUTO parameter,
computed in the R or r band (CFH12K
or Megacam).  This is an elliptical--aperture measurement, with the
aperture size based on the first moment of the object light
distribution, with an automatic correction for neighbour contamination
(we use the MASK\_TYPE CORRECT option).  This is the only total magnitude we compute directly.
Total magnitudes in other filters are always derived from these, using the
appropriate aperture colours.  

Finally, the catalogues are matched to the CNOC2 redshift catalogues
(including additional redshifts from our Magellan programme).  Although
the relative astrometry of each MOS field in the original CNOC2 survey
is good, the global astrometric solution for each patch has significant
offsets which makes matching sources difficult.  Therefore we first improve
upon the CNOC2 astrometry by re-solving each MOS pointing, using the
publicly available {\sc astrometry.net} software \citep{astrometrydotnet,lang09}.  This provides more accurate coordinates for
all CNOC2 galaxies, and we then match them with the closest source in
our new optical and infrared catalogues.

\subsubsection{K-corrections}
We compute k-corrections for all galaxies with redshifts, using the
{\sc kcorrect} v4.1.4 software of \citet{kcorrect}.  We use the 3\arcsec\ aperture
magnitudes, in $ugrizBVRIK$ where available.  We do not include the
Spitzer data here, as it is not critical for k-corrections and would
require us to use degraded (i.e. lower resolution) optical data for the
psf-matched photometry.  In the few cases where
$K_s$ imaging exists from both SOFI and INGRID, we take the SOFI value.  
For the optical photometry, we use the native systems and use the
appropriate filter and instrument response curves.  Since the $K_s$ data
were calibrated to 2MASS, we use the standard 2MASS response curves for
these fluxes (and convert to the AB system).  Magnitudes were corrected for Galactic extinction using
the \citet{SFD} maps, where we have taken a single typical value
for each CNOC2 patch\footnote{Specifically, we model the extinction as
  $A_\lambda = A_\circ-1.5\log_{10}\lambda/{\mu m}$, with $A_\circ=-1.415$ in the 14h
and 21h patches, $A_\circ=-1.285$ for the 2h patch, and $A_\circ=-1.8$ for
the 9h patch.}.  A systematic calibration
uncertainty was added in quadrature to the uncertainty computed by {\sc
  SExtractor}.  This was typically 0.02-0.03 for the optical filters,
but 0.05 in $u$ and $K_s$.  Finally, we only fit galaxies with coverage
in at least four different filters.  

For this paper we k-correct the magnitudes to $z=0.4$, to minimize the
size of this
correction.  No evolution correction is applied.  
Thus, throughout this paper we always quote magnitudes corrected to the
$z=0.4$ observed frame, indicated by a superscript; for example,
${^{0.4}r}$ represents the observed $r$ magnitude at $z=0.4$.

\begin{figure}
\leavevmode \epsfysize=8cm \epsfbox{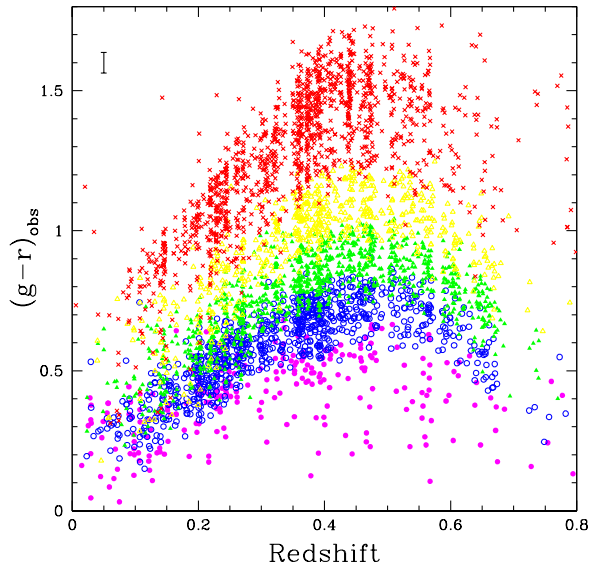} 
\caption{The observed Megacam (g-r) colours as a function of redshift,
  colour-coded according to their $z=0.4$ frame colours. The
  filled, magenta circles represent galaxies with ${^{0.4}(g-r)}<0.6$, and the
  blue, green, yellow and red symbols (open circles, filled and open
  triangles, and crosses, respectively) represent increasingly red bins in 
  $z=0.4$ colour, $0.2$ mag wide. The error-bar in the top-left corner
  shows the typical 1$\sigma$ uncertainty, including zeropoint error,
  for $r<22.5$ galaxies at $0.25<z<0.55$.\label{fig-grz} }
\end{figure}
Figure~\ref{fig-grz} shows the observed (g-r) colour as a function of redshift,
for all galaxies with k-corrections.  The points are colour-coded according
to their $z=0.4$ colour, as indicated.  The fact that the
points are properly segregated at all redshifts is one demonstration
that the k-corrections we compute are reliable (e.g., the galaxies with
the reddest $z=0.4$--frame colours are generally the reddest in observed
colour at other redshifts).  Although for some fits
the reduced $\chi^2$ value is large, we find no tendency for these
k-corrections to be distributed differently as a function of colour and
redshift than those with good fits.  Therefore, we only omit data where
there are fewer than four available filters, which would be of limited
use to us here anyway.

Absolute magnitudes are computed from extinction and k-corrected
magnitudes at $z=0.4$, assuming $\Omega_m=0.3$, $\Omega_\Lambda=0.7$
and $h=0.7$ as usual.

\subsubsection{Combining similar filters}\label{sec-filtercomb}
We will combine the photometry from different instruments, when the
bandpasses are similar.  By directly comparing galaxies in the
relevant magnitude range with both CFH12K and Megacam coverage
(approximately 200--350, depending on the filter), we compute the following
transformations (lower-case filters refer to Megacam, while upper-case
filters are CFH12K):
\begin{align}
g=&B+0.058-0.588(B-V),& \\
r =& R-0.07+0.261\left(V-R\right),& \\
i=&I+0.119& \text{if $(B-V)>1$,}\\
i=&I-0.11+0.2(B-V)& \text{if $(B-V)\leq1$.}
\end{align}
These transformations have an {\it rms} dispersion of 0.05 mag in $i$
and $r$, and 0.08 mag in $g$.
\begin{figure}
\leavevmode \epsfysize=8cm \epsfbox{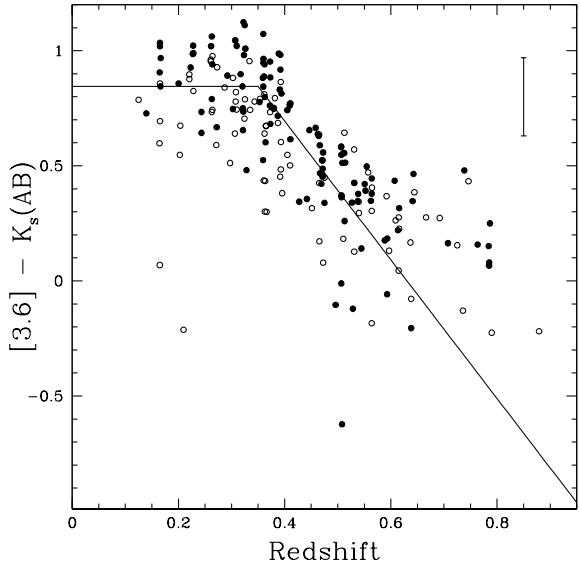} 
\caption{The observed [3.6]-$K_s$ colours as a function of redshift, for
  galaxies with $K_s<20$,  shows a
  fairly well-defined relation with rms scatter of about 0.22 mag.  The error-bar in the top-left corner
  shows the typical 1$\sigma$ uncertainty, including zeropoint error;
  this accounts for most of the scatter.
  There is little dependence on star formation history: the filled circles
  indicate galaxies with red colours, ${^{0.4}(g-r)}>1.2$.  The solid line
  shows a bilinear fit to this relation; we use this to convert [3.6] magnitudes
  to equivalent AB $K_s$ magnitudes at the appropriate redshift, then
  apply a k-correction to $z=0.4$, as in \citet{cnoc2_ir}.  This allows us to treat the
  ground-based and space-based near-infrared measurements together.\label{fig-s1tok}}
\end{figure}

In all cases, we will use the Megacam $gri$ magnitudes if available;
otherwise we will use the CFH12K $BRI$ magnitudes, appropriately
transformed using these relations.

It is also useful to combine the INGRID, SOFI and IRAC 3.6$\mu$ m
data into a single equivalent $K_s$ band magnitude.  This is trivial for
the INGRID and SOFI data, since they are both calibrated to the 2MASS
system.  Although the 3.6$\mu$ m data is obviously distinct from $K_s$,
the relative insensitivity of this part of the spectrum to spectral
type makes a simple conversion to $K_s$ magnitudes a sensible thing to do for our purposes.  We follow a
strategy similar to that outlined in \citet{cnoc2_ir}, noting that the
[3.6]-$K_s$ colour is primarily sensitive to redshift, as shown in
Figure~\ref{fig-s1tok}.  This relationship has little dependence on intrinsic
galaxy colour, and is similar whether we use SOFI or INGRID
magnitudes, demonstrating indirectly that there is no systematic shift between
their respective zeropoints.  We fit this relationship with a constant
value for 
$z<0.35$, and a straight line at higher redshift, as shown.  We use
this fit to convert all [3.6] micron magnitudes to an equivalent $K_s$
magnitude at that redshift, then apply the k-correction (to $z=0.4$)
computed for the $K_s$ filter.  Comparing galaxies with both ground-IR and
IRAC data, we find an rms dispersion in this equivalent $K_s(AB)$
magnitude of about 0.22 mag, which primarily reflects uncertainty in
the ground-based zeropoint (about 0.1 mag) and a small residual colour
term (also about 0.1 mag).  
The values of $K_s(AB)$ in this paper will always use the IRAC data
where available, otherwise the SOFI or INGRID (in that order).

\subsection{Group and field definition}
We use the group centres and velocity dispersions as described in
\citet{McGee}, based on a slight modification of the method in
\citet{CNOC2_groupsI}.  Group members are considered to be all galaxies
within 500 kpc of 
the luminosity--weighted centre, and within twice the group velocity
dispersion in redshift.  If there are two groups potentially satisfying
these criteria for a given galaxy, the galaxy is assigned to the group to
which it is closer spatially.  Typical groups include $\sim 10$
confirmed members, and velocity dispersions are formally determined to
within $\pm 100$km/s, using the gapper algorithm of \citet{Beers}.

\begin{figure}
\leavevmode \epsfysize=8cm \epsfbox{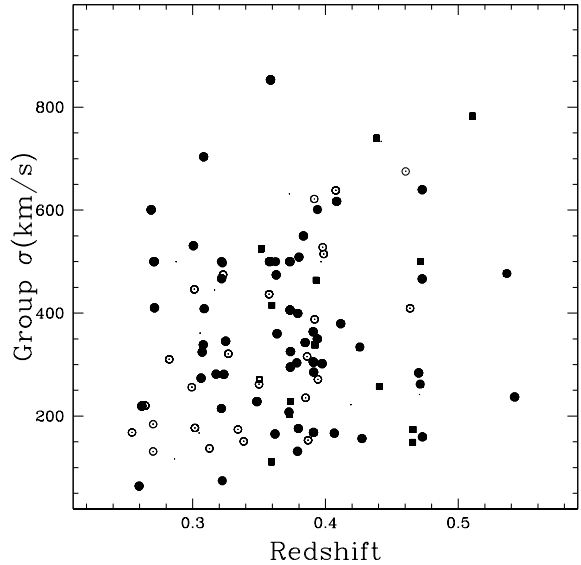} 
\caption{
The rest-frame velocity dispersion of all groups in our sample is shown
as a function of group redshift.  Error bars are not shown, for
clarity, but are typically $\pm 100$km/s.  Filled symbols represent
groups with near-infrared imaging, as described in \S~\ref{sec-nir}.
Squares are groups with {\it HST} coverage (\S~\ref{sec-morph}), while
circles are those without.  Small points are groups without new optical
imaging from CFHT, and are excluded from all analysis in this paper.
\label{fig-sig_dist}}
\end{figure}

Figure~\ref{fig-sig_dist} shows, as a function of group redshift, the distribution of $\sigma_{\rm
    rest}$,
the rest-frame velocity dispersion of the host group for each galaxy in
our sample with $0.25<z<0.55$.  These
velocity dispersions
are actually upper limits, as they include intrinsic dispersion due to
redshift uncertainties.  Most of
the galaxies come from groups with $\sigma<500$~km/s; in
\citet{cnoc2_ir} we show that the stellar masses and lensing masses of
these groups are also in good agreement with these velocity
dispersions.  Thus, our sample is dominated by true groups, and none of our
results are changed if we exclude groups with $\sigma>500$~km/s.  The
figure also shows which groups have optical, near-infrared and {\it
  HST} follow-up imaging, as described in the caption.  We see that the
subsample of groups considered in this paper is representative of our
full group sample.

We will define our field sample to be all galaxies in the survey that are not assigned to a
group, either using our current definition or with the more
encompassing friends-of-friends definition employed in previous
papers.  However, our group sample is not complete, partly because they
were selected from a sparsely--sampled redshift survey \citep{McGee}.
Therefore our field sample is more representative of the total galaxy
population; we expect that approximately half of all galaxies in this
sample are themselves in groups or clusters \citep{Eke-groups,McGee,McGee-accretion}.

\subsection{Statistical weights and Completeness}\label{sec-weights}
Our spectroscopic selection function is complex because certain fields
within the original CNOC2 survey area were retargetted with LDSS2 on
Magellan, improving both the completeness and depth.  Since the fields
were chosen to contain groups, the selection function also depends on
environment. In previous papers \citep[e.g.][]{cnoc2_ir,CNOC2_groupsI,cnoc2_irac} we
have implemented different, complex weighting schemes to try and
account for this as best as possible.  These methods are never perfect,
and rarely have any substantial influence on the results.

Here we adopt a slightly simpler scheme, and apply weights that depend
only on the $R$ magnitude (from the original MOS imaging on which the
spectroscopic targets were selected) and their location inside or
outside of our Magellan target fields.  That is, for the regions
observed at Magellan, we compute completeness as a function of $R$, for
all galaxies with spectra in that field (including those from the
original CNOC2 survey).  We do the same for the regions not covered by Magellan.
This then provides a statistically-complete sample to
$R<21.5$; the Magellan fields are statistically complete to about 0.5
mag fainter than this.

Our photometry is all much deeper than our spectroscopic limit.  We
would like to consider galaxy properties as a function of the
reddest-possible band, which provides the best tracer of stellar mass.
Ideally this would be the $K_s$ and [3.6]$\mu m$ band data; however, we
have a larger sample if we use $i$, so in this paper we will present
our data as a function of both magnitudes.
In our redshift range of interest,
$0.25<z<0.55$, the reddest galaxies have $(r-i)\sim 0.8$ and
$(r-K_s)\sim 2.5$, so the 
spectroscopic limit $R=21.5$ corresponds to $i\sim 20.7$ and
$K_s\sim 19$.  In
$z=0.4$ frame absolute magnitudes, at $0.25<z<0.55$, these limits are
$-19.8>{^{0.4}M_i}>-21.8$ and $-21.5>{^{0.4}M_{K_s}}>-23.5$.  We use a $1/V_{\rm max}$
weight to construct a statistically complete volume--limited sample within this
redshift range, with ${^{0.4}M_i}<-19.8$ and ${^{0.4}M_{K_s}}<-21.5$.
Note that this volume weight gets very large (greater
than 20) within about 0.2 mag of these limits, so we
restrict most of our analysis to ${^{0.4}M_i}<-20.0$ or ${^{0.4}M_{K_s}}<-21.7$.

\subsection{Measures of activity}\label{sec-act}
In addition to the colours of interest here, we consider other measures
of galaxy ``activity''.  This includes the rest-frame equivalent width
of the [OII] emission line, $W_\circ([OII])$.  These were measured on
all spectra as described in \citet{Whitaker}, using the same algorithm
as that described in \citet{B+97}.
We take $W_\circ([OII])>10$\AA\ as a
secure indication of activity (but do not distinguish between star formation and AGN).

We also use the k-corrected IRAC [8]$-$[3.6] micron colours in 5\arcsec\ apertures, where available, as a more
sensitive probe of activity, as described in \citet{cnoc2_irac}.   \citet{cnoc2_irac}
show that $f_8/f_{3.6}>0.5$ is a good indicator of activity, since galaxies
with brighter 8$\mu m$ emission at this redshift tend to be dominated
by PAH emission lines related to star formation, or continuum emission
from AGN.  We will
adopt the same definition for passively-evolving galaxies here,
$f_8/f_{3.6}>0.5$ here.  Although we measure the photometry somewhat
differently from \citet{cnoc2_irac}, this does not make a large enough difference to
motivate a change in threshold.  

\subsection{Morphology}\label{sec-morph}
We use the galaxy morphologies presented in \citet{Wilman_morph}.
The morphologies are based on {\it HST} ACS observations of a subset of our groups.
The classifications onto the Revised Hubble Scheme were made independently by two of us (AO and JM).  To simplify
somewhat, we group E and E/S0 together into a single ``E'' class.
Similarly, S0/E, S0 and S0/a will all be treated as ``S0''.  Sa and Sb
will be termed ``early S'', while Sc, Sd and Sm will be ``late S''.  Any
irregular galaxy, or those identified as merging or interacting,
will all be considered ``Irr/M''.  For these purposes, notes of
peculiarity will be ignored.

\subsection{Stellar mass}\label{sec-smass}
Previously \citep{cnoc2_ir,cnoc2_irac} we presented stellar mass
estimates for a subset of galaxies in the present sample, based on simple
stellar population models, and we will soon publish more accurate
values computed from fits to the full spectral energy distribution available
to us, including GALEX data (McGee et al. in prep).
In this paper however, we aim to compare the properties of group and field
galaxies as directly as possible, thus restricting our analysis to quantities closely related to
observables (i.e., instrumental colours and magnitudes at
$z=0.4$).  
Nonetheless it is
useful to be able to relate absolute magnitudes in the $z=0.4$ frame
colours to an approximate stellar mass for reference.  For this
purpose, we compute two
\citet{BC03} models galaxies with a \citet{Chab}
initial mass function and metallicity $0.2Z_\odot$, as expected to be
appropriate near the limit of our data.  The first is a dust-free, single stellar population of age
8.0 Gyr (the time between $z=5$ and $z=0.4$), and the other assumes a
constant star formation rate for the same duration, and includes 1
magnitude of dust extinction.  
 In the first case, we find
$M/{^{0.4}L_{K_s}}=0.35$ and $M/{^{0.4}L_i}=1.6$, while for the second
(younger) model
$M/{^{0.4}L_{K_s}}=0.22$ and $M/{^{0.4}L_i}=0.78$.  Thus, star-forming
galaxies are 0.8 mag brighter in $I$, and $0.5$ mag brigher in $K_s$,
than old galaxies with the same stellar mass.


Recall that our sample is statistically complete, for galaxies of all colour,
at $^{0.4}M_i<-20$ and $^{0.4}M_{K_s}<-21.7$.  Using the simple models
above, our sample is complete, for all galaxy types with
$M\gtrsim 1\times 10^{10} M_\odot$ ($I$-selected), or $M\gtrsim 2\times 10^{10} M_\odot$ ($K_s$-selected).  Note that the bluest objects in
our sample reach luminosities as faint as $^{0.4}M_{K_s}=-19$, which
corresponds to a stellar mass of $M=1.0\times10^{9}M_\odot$.  It will be
interesting to compare the field and group populations of such
low-mass, star forming galaxies --- even though the corresponding old
population is not represented.  

\section{Results}\label{sec-results}
\subsection{Colour distributions}
We begin by considering colours that span the age-sensitive 4000\AA\
break.  At $z=0.4$ (to which all our magnitudes are k-corrected), this
is $(B-R)$ in the CFH12K images, and $(g-r)$ for Megacam.  To make the maximum use of our data,
we transform CFH12K magnitudes to the Megacam system, as described in
\S~\ref{sec-filtercomb}.  There are 2215 galaxies in our redshift
range with spectroscopy and good g, r and i magnitudes.  Of these, 333 are group
members.

\begin{figure}
\leavevmode \epsfysize=8cm \epsfbox{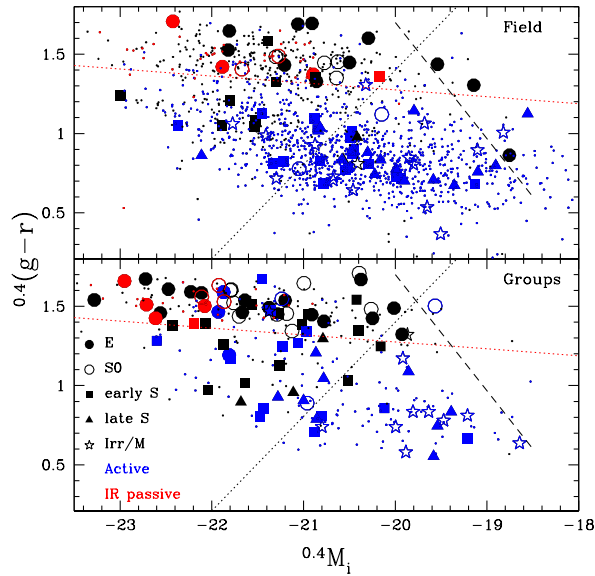} 
\caption{A colour-magnitude diagram for group galaxies (bottom panel)
  and field galaxies (top panel), in our $0.25<z<0.55$ sample.  The
  large symbol types are keyed to galaxy morphology as described in
  \S~\ref{sec-morph} and indicated in the legend; small circles have no
  ACS coverage, hence no morphological information.  Blue symbols
  represent those galaxies that show either strong [OII] emission, or have an IR
  excess, indicating star formation or AGN activity.  Red symbols are
  galaxies that show no IR excess.  Black symbols are galaxies with no
  {\it Spitzer} data to measure an IR excess, and weak, absent or
  unmeasured [OII] emission.  The black, dashed line shows our
  colour-dependent completeness limit at $z=0.25$.  Brighter than this there is
 absolute magnitude incompleteness, which we can correct with a $V_{\rm max}$
  weight.  Fainter than ${^{0.4}M_i}=-19.8$ the sample is
  incomplete for red galaxies.  The black dotted line represents
  approximately a constant mass limit of $1\times10^{10}M_\odot$,
  computed by simply joining the young and old population synthesis
  models described in the text.  The red dotted lines
  represent our division between red and blue galaxies, based on a 0.15
  mag offset from a fit to the red sequence.\label{fig-gri} }
\end{figure}

In Figure~\ref{fig-gri} we show the $^{0.4}(g-r)$ colours as a function of $^{0.4}{M_i}$, for group and field members in our
sample.  The point symbols are keyed to galaxy morphology, for those
with ACS imaging, as indicated.  
As expected, the elliptical galaxies follow a
tight red sequence, in both environments, with few exceptions.  This is
generally true also of the S0 galaxies, though a few of these show blue
colours.  There are almost no late-type spirals on the red sequence.
\begin{figure}
\leavevmode \epsfysize=8cm \epsfbox{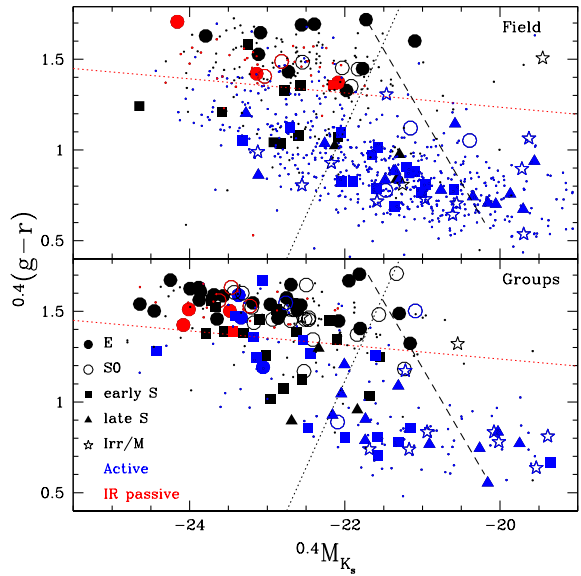} 
\caption{The same as Figure~\ref{fig-gri}, but as a function of
  absolute $K_s$ magnitude.  Here, $K_s$ is converted from {\it Spitzer}
  [3.6] micron fluxes if available; otherwise we use our SOFI and
  INGRID data.  The black dotted line represents
  approximately a constant mass limit of $2\times10^{10}M_\odot$,
  computed by simply joining the young and old population synthesis
  models described in the text.  
\label{fig-grk}}
\end{figure}

The data are colour-coded according to their level of activity. Blue
symbols are galaxies with either strong [OII] emission, or an infrared excess, as
described in \S~\ref{sec-act}.  To be conservative, red-coloured points
are only those for which IRAC colours are available, and for which
there is no IR excess.  These are likely to be truly passive galaxies;
we can't say the same about galaxies with weak or absent [OII], since
our spectra have relatively low S/N and this line is relatively weakly sensitive
to star formation.

Our best tracer of stellar mass is the near-infrared $K_s$ magnitude,
converted from {\it Spitzer} [3.6] magnitude if available, and
otherwise taken from our SOFI and INGRID data.  This results in a
smaller sample of only 1429 galaxies (303 in groups) but is the
least-sensitive to stellar population effects.   
Because of the weak colour-dependence of $M/^{0.4}{L_{K_s}}$, a 0.5 mag
interval in $^{0.4}{M_{K_s}}$ corresponds to only a factor $~\sim 2$
variation in stellar mass; thus our $K_s$--selected sample is
reasonably close to a mass--selected sample. 
We show the $^{0.4}(g-r)$
colours as a function of $^{0.4}{M_{K_s}}$ in
Figure~\ref{fig-grk}.  The same features are visible here, as in
Figure~\ref{fig-gri}; in particular,   
the majority of blue galaxies in our sample
are low-mass, well below our colour-completeness limit of $\sim 2\times
10^{10}M_\odot$, shown as the dotted black line in the figure. 
We emphasize that above this limit we have
a statistically complete sample of group and field galaxies,
independent of their colour.  This should be compared with the higher redshift groups
in the DEEP2 survey \citep[e.g.][]{DEEPII_envt_again}, for which only
the very most massive red galaxies are detected.

\begin{figure}
\leavevmode \epsfysize=8cm \epsfbox{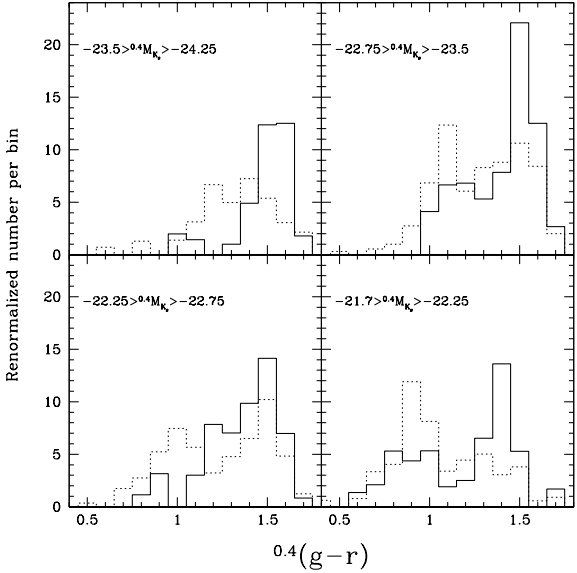} 
\caption{The colour distribution of galaxies in four
  near infrared luminosity bins, as indicated.  Field galaxies are
  shown in dashed histograms,
  while group galaxies are shown with solid lines.  Distributions are weighted by
  statistical sampling weights and $1/V_{\rm max}$, but then
  renormalized to reflect the total number of {\it group} galaxies in
  each luminosity bin shown.  
\label{fig-grkdist}}
\end{figure}
In  Figure~\ref{fig-grkdist} we show the $^{0.4}(g-r)$ colour
distribution in four bins of infrared luminosity.  Adopting an
approximate $M/^{0.4}L_{K_s}=0.3$, intermediate between the young and old models
described in \S~\ref{sec-smass}, these bins are centred approximately on 
stellar masses of $1.2\times10^{11}M_\odot$, $6.3\times10^{10}M_\odot$,
$3.5\times10^{10}M_\odot$ and $2.2\times10^{10}M_\odot$.
The galaxies are
weighted as described in \S~\ref{sec-weights}, including the $V_{\rm
  max}$ weight.  However, the group-member histograms are renormalized
to the total number of galaxies in each luminosity bin, so the y-axis
is more reflective of the unweighted sample size.  The field histograms are
normalized to match the number of group-galaxies in each bin, so the
shapes can be directly compared.

This figure shows the now well-established bimodal distribution of
galaxy colours.  At all luminosities, the group sample
shows more galaxies in the red part of the distribution than the
corresponding field sample.  Apart from that, the shape of the colour
distribution looks remarkably similar for both environments.  This also
holds for lower-masses, where the sample is incomplete for red galaxies.
The colour distribution of faint ($^{0.4}M_{K_s}>-21$) galaxies along the blue sequence in the
group sample is indistinguishable from that
in the field, as apparent from Figure~\ref{fig-grk} and statistically
confirmed with a K-S test.

\begin{figure}
\leavevmode \epsfysize=8cm \epsfbox{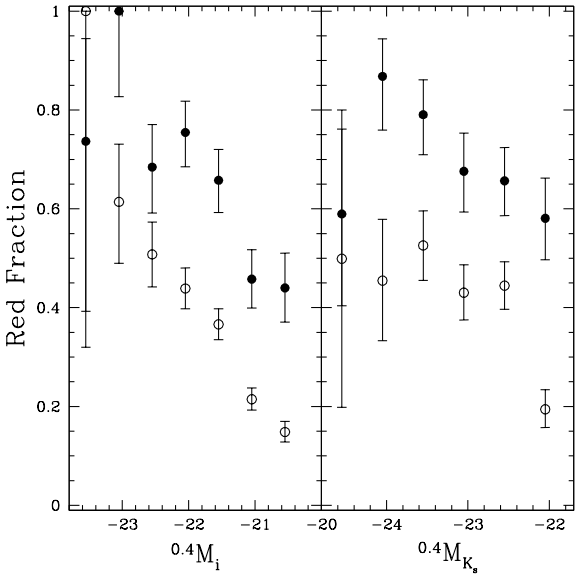} 
\caption{The fraction of red galaxies as a function of absolute
  magnitude, in the group (filled circles) and field (open circles) samples.  Red galaxies
  are defined as those no more than 0.15 mag bluer than a fit to the
  red sequence in $^{0.4}(g-r)$, as shown in Figures~\ref{fig-gri} and \ref{fig-grk}.  There is
  a significant excess of red galaxies in the group sample, at
  fixed luminosity.
\label{fig-redfrac}}
\end{figure}

Thus, as at low redshift, the dominant environmental effect observed is
an increase in the fraction of red galaxies in dense environments.  To
quantify this, we fit a line to the red sequence and define a colour
cut 0.15 mag bluer than this limit, as shown in
Figures~\ref{fig-gri} and \ref{fig-grk}.  We then compute the red fraction as a function of
magnitude, weighting for incompleteness and volume effects as usual.
The result is shown in Figure~\ref{fig-redfrac}; error bars on weighted
fractions, here and throughout the paper, are computed using binomial
statistics \citep{Gehrels}.
At fixed luminosity, an additional $\sim 20\pm 7$ per cent of the group
population is found on the red sequence, compared with the field; this
is true whether we use $i$ or $K_s$ luminosities, though the trend is
weaker in the latter case.  The well-known trend for lower-mass galaxies to have
higher blue fractions \citep[e.g.][]{BaldryV} holds equally well for
our group galaxies. However, the key point here is that the galaxy
population has a clear dependence on environment, independent
of any trends with stellar mass.  
\begin{figure}
\leavevmode \epsfysize=8cm \epsfbox{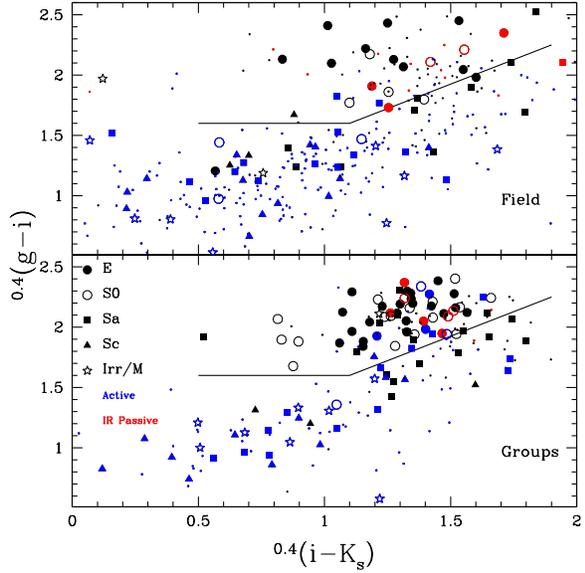} 
\caption{A colour-colour diagram for group (bottom) and field (top)
  galaxy populations.  The y-axis is mostly sensitive to age, while the
  x-axis is more sensitive to dust.  The passive, early-type population
  is well-separated from the more normal, active population, although
  they overlap in $(g-i)$ colour.  The solid lines show our choice for
  isolating the truly passive population, which lie to the left of
  these lines.
\label{fig-giik}}
\end{figure}

\subsection{Dusty red galaxies}\label{sec-dustyred}
The red sequence is known to include some actively star-forming
galaxies that are reddened by dust \citep[e.g.][]{C17-dust}.
At low redshift, these tend to be mostly early-type spiral galaxies
with low star-formation rates, or highly-inclined normal spirals \citep{STAGES-dust}.
In Figure~\ref{fig-giik} we use a colour-colour diagram, with
rest-frame colours chosen to be comparable to those in \citet{C17-dust},  to try and separate these populations. At $z=0.4$, $(g-i)$ is an age-sensitive feature
spanning the 4000\AA\ break, while $(i-K_s)$ is more sensitive to dust.
The field population in Figure~\ref{fig-giik} shows a clear extension of the blue sequence
toward red $(g-i)$ colours, but offset in $(i-K_s)$ from the population of E
and S0 galaxies, which show little or no activity.  The solid lines
show an approximate colour cut which isolates the population expected
to be mostly inactive, from the more normally star-forming population.
The same division is evident in the group sample, but now the
inactive population is much more prominent.

We therefore make our best attempt to identify galaxies which appear to
be truly inactive.  Our primary requirement for a passive galaxy is one that
not only lies on the $^{0.4}(g-r)$ red sequence defined above, but also
shows colours typical of passively evolving galaxies in the $^{0.4}(g-i)$--$^{0.4}(i-K_s)$ diagram (Figure~\ref{fig-giik}).
We also remove from that sample any galaxy with strong [OII] emission
($W_\circ(O[II])>10$\AA), or an IR excess as measured from available {\it
  Spitzer} data (Section~\ref{sec-act}).  
This definition removes about 20 per cent of galaxies from the
$^{0.4}(g-r)$ red sequence, approximately independent of magnitude and
environment.  Most of this difference is due to the colour-selection;
the additional contribution from [OII]--emitting or IR excess galaxies
is relatively small.  

\begin{figure}
\leavevmode \epsfysize=8cm \epsfbox{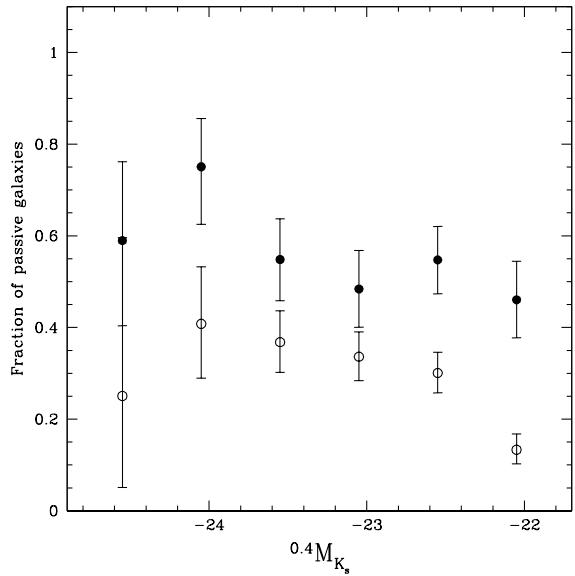} 
\caption{
The fraction of galaxies identified as having no ongoing star
formation from their $^{0.4}(g-i)$--$^{0.4}(i-K_s)$
  colours as shown in Figure~\ref{fig-giik}, and an absence of strong
  [OII] emission or an IR excess, if either of these indicators is
  available.   Group galaxies are shown with filled circles, and the field with
  open circles.   The field is a global sample, of which
  approximately 50 per cent are themselves in groups or clusters.  Our
  data are consistent with isolated (or ungrouped) galaxies having few
  or no inactive galaxies whatsoever (see text for more details).
\label{fig-dustyred}}
\end{figure}

Our best estimate of the truly inactive fraction is shown
in Figure~\ref{fig-dustyred}.  The most important result here is that
inactive galaxies are significantly more common in the group
environment, at all magnitudes.  Thus, the higher incidence of group
galaxies on the $(g-r)$ red sequence is not simply reflecting a dustier
population.  Secondly, in both environments, the fraction of inactive
galaxies is only weakly dependent on  $M_{K_s}$ luminosity (closely
related to stellar mass), though the large error bars on the most
luminous points makes it difficult to draw strong conclusions. 

Recall finally that our field sample consists of galaxies in all
environments, including groups.  In particular, approximately 40 per cent of our
lowest-mass field galaxies, and 60 per cent of our highest mass field galaxies, are
expected to be in groups or clusters \citep{McGee-accretion}.  Using this, the fraction
of ``ungrouped'', inactive galaxies can be estimated from
Figure~\ref{fig-dustyred}.  For example, at bright luminosities, $54\pm6$ per
cent of group galaxies and $35\pm3$ per cent of field galaxies are inactive.
Since $\sim 55\pm 5$ per cent of the field galaxies themselves are expected to be
in groups, this leaves room for only $\sim 11\pm15$ per cent inactive
galaxies among the isolated component of our field sample.  This is consistent with zero, for
all luminosities considered here, suggesting
that almost {\it all} of the ``red-and-dead'' galaxies found in field
surveys at this redshift are associated with dense environments.

\subsection{Population Dynamics}
\begin{figure}
\leavevmode \epsfysize=8cm \epsfbox{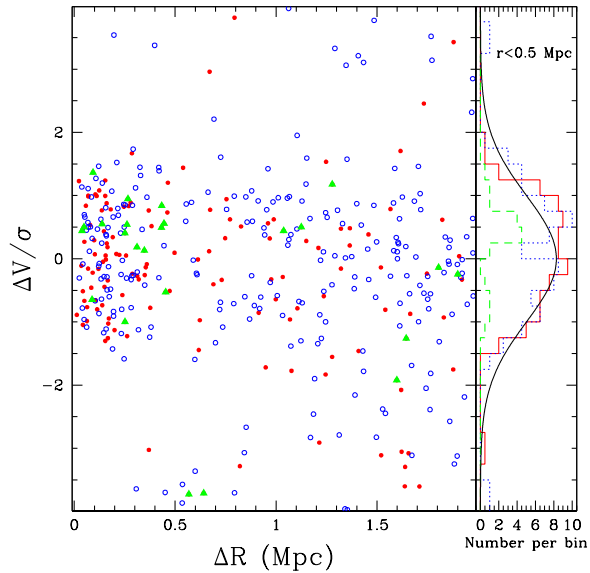} 
\caption{The rest-frame velocity offset, normalized to the group
  velocity dispersion, is shown as a function of physical distance from
  the group centre, for all galaxies with ${^{0.4}M_i}<-20.0$, near
  the groups in our sample.  Galaxies are divided into 
  red and blue populations (red, filled circles and blue, open circles,
  respectively) based on their $gri$ photometry, as shown in
  Figure~\ref{fig-gri}.  Galaxies on the red sequence, but with
  evidence of activity, as discussed in \S~\ref{sec-dustyred}, are
  indicated with green triangles.  
  The histograms on the right show the
  velocity distribution of each galaxy population, within $0.5$ Mpc.
  The smooth, black curve is a Gaussian with unit variance and zero
  mean, normalized to match that of the observed, red galaxy
  population.  Both the red (solid histogram) and the blue (dotted
  histogram) population are observed to
  have similar spatial and velocity distributions, somewhat flatter
  than a Gaussian.  The ``dusty
  red'' galaxies (dashed histogram) shows a double-peaked
  distribution characteristic of infall.
\label{fig-drdv}}
\end{figure}
We consider the dynamics of the different populations of group members,
in Figure~\ref{fig-drdv}.  This figure shows the velocity offset of every
group member brighter than ${^{0.4}M_i}=-20.0$, relative to the velocity dispersion of the host group, as
a function of distance from the group centre.  Galaxies are divided
into red-sequence and blue-cloud populations, based on their
$^{0.4}(g-r)$ colours (Figure~\ref{fig-gri}); green points denote
red-sequence galaxies with some evidence of ongoing activity as
described in \S~\ref{sec-dustyred}.  Overall, the group population
appears well separated from the surrounding field in velocity space,
and evidence for overdense structure can be seen well beyond 1 Mpc
(recall that all the analysis in this paper is restricted to the
population within 0.5 Mpc).  Secondly, the red and blue populations
have very similar phase space distributions.  Within 0.5 Mpc both
appear to have a somewhat flatter velocity distribution than a
Gaussian, although a Gaussian model cannot be ruled out with a 
Kolmogorov--Smirnov test.  The similar dynamics of the two populations
suggests that the blue galaxies are not preferentially contaminated by fore- and background
galaxies.  

Intriguingly, the red-sequence galaxies with evidence for
ongoing star formation do exhibit some dynamical differences from the
rest of the population.  Those found within $0.5$ Mpc have a velocity
distribution statistically inconsistent (0.6\% probability) with a
Gaussian with a mean and dispersion given by the main galaxy
population.  The double-peaked distribution is characteristic of an
infalling population; this may then represent a population in which
star formation is being truncated upon first accretion into a group
environment.  Whether or not this truncation is accompanied by a
 heavily obscured burst of star formation or not is unclear; we are
 currently analysing 24$\mu$m data from {\it Spitzer} that should shed
 light on this issue (Tyler et al. in prep).

\section{Discussion}\label{sec-discuss}
We have shown that the optical colour distribution of group galaxies
at $z\sim 0.4$ is very similar to that of the surrounding field, but with a
more prominent population of red galaxies. 
 While we cannot rule out a contribution
  from short-lived starbursts, or heavily dust-obscured star formation, 
we find no evidence that dense environments at this epoch are doing anything to
{\it induce} star formation on a large scale.  
This is particularly relevant, since
small groups are anticipated to be the prime environment for galaxy
interactions to take place, and these interactions might lead to
enhanced star formation.
Furthermore, as galaxies in groups are
destined to make up a substantial fraction of the massive cluster galaxy
population by $z=0$ \citep[e.g.][]{Berr+08,McGee-accretion}, it seems unlikely
that the red population in clusters was produced via such
starburst--inducing interactions within groups.  
\begin{figure}
\leavevmode \epsfysize=8cm \epsfbox{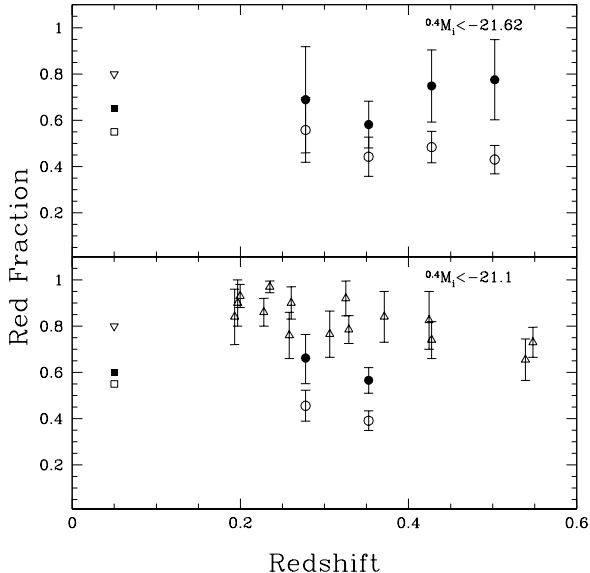} 
\caption{{\bf Bottom panel: }The fraction of red galaxies brighter than
  ${^{0.4}M_i}=-21.1$, in our group (filled circles) and field
  (open circles) samples as a function of redshift.  This is compared with the
  red fractions in X-ray luminous clusters over a similar redshift and luminosity
  range, shown as the open triangles, taken from \citet{Erica}. We also
  show an estimate of the red fraction locally, from the SDSS sample of
  \citet{Weinmann_1+06}.  The inverted triangles, filled and open squares
  represent clusters, groups and field galaxies, respectively.  {\bf
    Top panel:} Our full group and field sample, for a brighter luminosity limit of
  ${^{0.4}M_i}=-21.62$.  The brighter limit allows us to plot
  consistent measurements for our sample over the full redshift range shown.
\label{fig-redfracz}}
\end{figure}

It is therefore useful to compare the group population with that of
more massive galaxy clusters at a similar redshift. The most comparable
survey is that of \citet{Erica}, who 
  have measured the blue fraction for fifteen massive clusters at
  $0.3<z<0.5$, considering all galaxies within $R_{200}$
  brighter than ${^{k,e}M_r}=-20-5\log{h}=-20.77$. At $z=0.4$, the observed $i$ magnitude corresponds
  very closely to rest-frame $r$, and is only about 0.1 mag fainter on
  average. We find our $z=0.4$ $i$ magnitudes, uncorrected for
  evolution, correspond to the magnitudes in \citet{Erica}
  approximately as
\begin{equation}
{^{0.4}M_i}\approx {^{k,e}M_r}-0.3 \pm 0.2
\end{equation}
Thus, to compare with the sample of \citet{Erica}, we calculate the red fractions for
galaxies brighter than ${^{0.4}M_i}=-21.1$, which corresponds
  approximately to a stellar mass limit of $2.8\times10^{10}M_\odot$.
Our sample is only
colour-complete to this limit for $z\leq 0.39$, so we restrict the
analysis to these groups, and note that the clusters are probably also
somewhat incomplete in red galaxies above this redshift, as the \citet{Erica} sample has a very
similar selection function to ours.  We do not know the masses of the
groups well enough to estimate a robust value of $R_{200}$, but our
radial selection of $r<0.5$Mpc is designed to be approximately this
radius for the typical group in our sample; thus the comparison with
the cluster data is fairly well-matched.
The results are shown in
Figure~\ref{fig-redfracz}. The 
groups lie in an intermediate region of the diagram, between the field and the clusters.  If we interpret the groups as an
intermediate step in the hierarchical growth of clusters, this suggests
that, with time, galaxies in dense environments slowly migrate onto the
red sequence, perhaps via a stage where there is non-negligible star
formation still occurring as inferred by \citet{STAGES-dust}.

We will defer a precise comparison with groups and clusters from
different samples, over a wider range of redshifts, to a future paper
in which we present robust stellar mass estimates based on SED-fitting
over the full spectrum from {\it GALEX-}UV to {\it Spitzer}-IR.
However, in the meantime it is useful to make a crude comparison with
luminosity--limited samples in the low-redshift
Universe, and it is most straightforward to compare with the analysis
of \citet{Weinmann_1+06}, based on the \citet{Y07} group catalogue from the
SDSS.  
Using the above relation between ${^{0.4}M_i}$ and ${^{k,e}M_r}$, we estimate the
corresponding red fraction in clusters (haloes with
$M>10^{14}M_\odot$), groups 
($10^{13}>M/M_\odot>10^{14}$) and the field 
(all galaxies).  We include both the ``early'' and
``intermediate'' classifications of \citet{Weinmann_1+06} in the red fraction, since these populations
largely overlap in colour.  These local red fractions are plotted as
inverted triangles, filled squares, and open squares, respectively.  
Comparing the clusters with those of \citet{Erica}, we see the
well-known result that, at these relatively high stellar luminosities, there
is little evolution in the cluster red fraction
\citep[e.g.][]{Erica,Nakata,deP_BOK}.  
The same lack of evolution holds for the groups and even the field,
which itself is dominated by galaxies within groups.   This result is
also in agreement with \citet{BB04}, who find that for groups in the SDSS with $\sigma<500$~km/s,
the red fraction of galaxies with $M_r<-20$ is about
65\%, very similar to the value obtained from the \citet{Weinmann_1+06}
analysis. 
This lack of evolution doesn't
come as a complete surprise, as evolution since $z=1$ is
largely dominated by the low--mass population, with $M\lesssim 3\times
10^{10}M_\odot$ \citep[e.g.][]{Bell+07}.  


\subsection{Comparison with models}
 It has now been shown in several places that many simple models
  for galaxy formation predict too many faint, red satellite
  galaxies at low redshift.  For example, \citet{Weinmann_1+06} compare the SDSS data
  with model predictions from \citet{Croton05}, and show that real groups and
  clusters have a large population of faint, blue satellites that are
  almost completely absent in the models.  This appears to be the
  result of an oversimplified gas-stripping model, whereby all hot gas
  is immediately removed from satellite galaxies upon merging with a
  larger halo.  Recently, \citet{Font+08} have explored a more
  realistic model, based on the calculations of \citet{Ian_rps}.  In
  this model, satellite galaxies are able to retain some fraction of
  their gas following accretion, depending on their orbit.  

To compare with our data, we select field and group galaxies at $z=0.4$ from both the
\citet{Bower05} and \citet{Font+08} models, using the algorithm described in \citet{McGee}
to closely reproduce our observational selection criteria.  In
particular, group membership is determined from projected distances and
line-of-sight velocities using the same friends-of-friends algorithm
used to define the observed groups.  We use the
$z=0.4$ observed-frame SDSS $(g-r)$ colours and 2MASS $K$ magnitude predicted
by these models, to closely match the observed quantities we have
presented here.  In addition to the differences in gas-stripping
physics, the Font et al. model has several other different parameter
choices compared with the previous Bower et al. model, but these are of
limited interest here.
Our goal is to see how well the models reproduce the overall shape
of the colour distribution, and we neglect 
statistical uncertainties on the magnitudes, which are relatively small.

\begin{figure}
\leavevmode \epsfysize=8cm 
\leavevmode \epsfysize=8cm \epsfbox{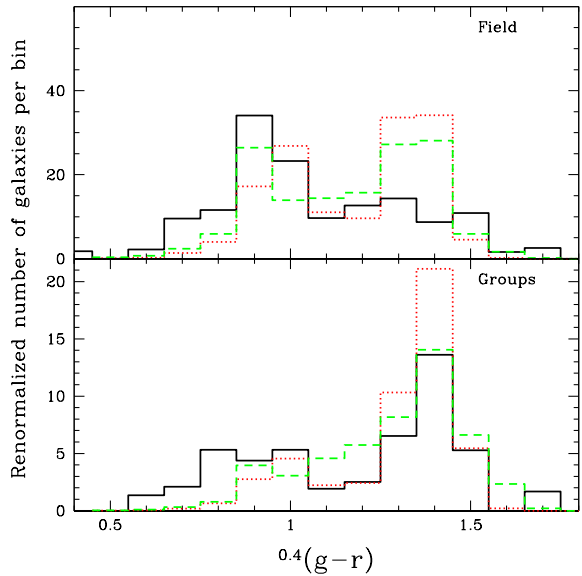} 
\caption{{\it Top panel: }The solid histograms show the observed $^{0.4}(g-r)$ colour distribution of galaxies in a
  narrow range of infrared luminosity, $-22.25>{^{0.4}M_{K_s}}>-21.7$, for
  our field sample.  This is compared with the predicted colours from
  the models of \citet[][red, dotted]{Bower05} and \citet[][green,
  dashed]{Font+08}, where the group and field environments are defined
  in a way that is analogous to our observational selection.  The observed distributions are statistically
  weighted, and all histograms are renormalized so their integrals reflect the
  total number of observed galaxies in the appropriate sample. 
{{\it Bottom panel:}  The same, but for our group
    samples.  
\label{fig-font}}}
\end{figure}
\begin{figure}
\leavevmode \epsfysize=8cm \epsfbox{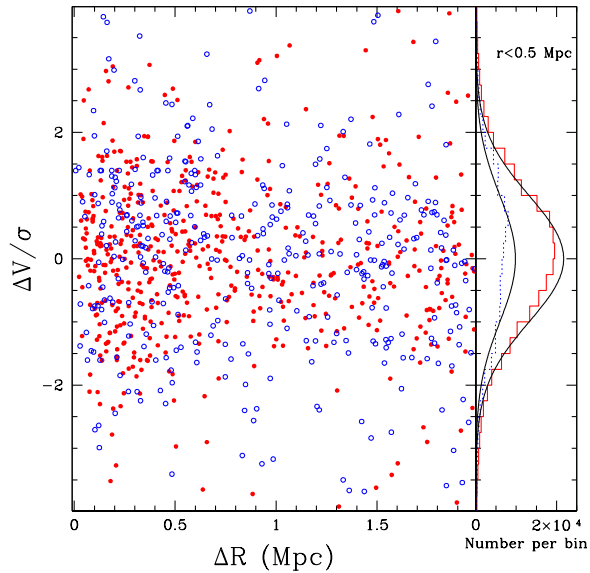}
\caption{The equivalent of Figure~\ref{fig-drdv} for the
  \citet{Font+08} model, this figure shows the velocity distribution as a function
  of projected distance for a sample of galaxies with ${^{0.4}M_i<-20}$
  in the mock groups. The red points and histogram correspond to
  galaxies with ${^{0.4}}(g-r)>1.2$, while the blue points represent
  bluer galaxies.  In the left panel, only a random 0.13 per cent of
  the model predictions are shown; this yields a sample size comparable
  to that of our observations.  In the right panel, Gaussian models with unit
  variance and zero mean are overplotted, normalized to the total area
  of the red and blue galaxy populations, for comparison.  Both
  distributions are statistically distinct from Gaussians, and from
  each other.  However, this distinction cannot be made if the sample
  size is reduced to match that of our observations.
\label{fig-modeldyn}}
\end{figure}

In Figure~\ref{fig-font},  we show the colour histograms for both models, compared with our
data.  We focus only on the lowest luminosity galaxies for which our
data are complete, $-22.25< {^{0.4}M_k}<-21.7$, since these are the most
sensitive to environmental effects. Both models do a reasonable job at
reproducing the observed bimodal colour distribution, but overpredict
the fraction of galaxies in the red peak.  Specifically, if we define
red galaxies as those with $^{0.4}(g-r)>1.2$ then the predicted
red fraction in the field is 54\% and 50\% in the Bower and Font models,
respectively, while the observed value is only $28\pm 2$\%.  Note that this
discrepancy is not due to a mismatch in the near-infrared luminosity;
the predicted red fractions change by only $\sim 5$ per cent within
$\pm 0.5$ mag of the luminosity range shown here.

Red, low-mass galaxies in these models are almost exclusively satellite
galaxies for which the gas supply is assumed to have been lost.  Thus,
the group environment is expected to be the key that controls the
overall fraction of red galaxies. 
In the bottom panel of Figure~\ref{fig-font} we actually see a
remarkably good qualitative agreement between the models and the observed colour
distribution in groups.  The older, Bower model predicts
a red fraction 75\%, still much too large compared with the observed
$56\pm 8$\%.  The Font model achieves its goal of reducing this number, to
67\%; however, it does so by overpopulating the ``green
valley'' between the red and blue sequences.  This is likely a
consequence of adopting a slow strangulation timescale for every
satellite galaxy; a better match might be achieved by only truncating star formation in a
subset of galaxies, for example based on their orbital parameters.
We will explore such an adjustment in a future paper.

Interestingly, both models predict that almost all of 
these blue ``group'' galaxies are found in haloes with $M<1\times10^{12}M_\odot$;
that is, they are nearby, but not actually incorporated into the main
group halo.  We therefore explore the dynamics of the galaxies in the Font model\footnote{The results are similar if
we consider the \citet{Bower05} model instead.}
in Figure~\ref{fig-modeldyn}.  This includes all galaxies with
${^{0.4}M_i}<-20$, and so can be compared directly with
the observations from Figure~\ref{fig-drdv}.  We first note that the
blue population, despite arising from galaxies outside the main dark
matter halo that defines the group, are clearly dynamically linked to
the group.  Therefore they are physically associated with the group,
and not randomly projected field galaxies.  Secondly, the distributions of both
the red and blue populations are statistically inconsistent with a
Gaussian, showing significantly flatter peaks.  They are also
statistically distinct from one another.  However the difference is
small, and if we choose a
random subset of the model galaxies, so the sample size is comparable
to that of our observational dataset, we can no longer statistically
distinguish these distributions from a Gaussian.  Therefore much larger
data samples will be required to make use of these predictions.

\section{Conclusions}\label{sec-conc}
We have presented an analysis of the optical and near-infrared
properties of galaxies in small groups at $z=0.4$.  We measure colours
consistently, using aperture photometry on {\it psf}-matched images, and
minimize k-corrections by analysing the observations in the $z=0.4$
frame.  We use the optical $^{0.4}(g-r)$ colours, which bracket
the 4000\AA\ break at this redshift, as our primary probe of the
stellar population.  Following \citet{C17-dust} and \citet{cnoc2_irac}
we also make use of the $i$ photometry and {\it Spitzer} IRAC
photometry, respectively, to distinguish dusty, star-forming galaxies
from truly passive galaxies with red $^{0.4}(g-r)$ colours.
Total luminosities are computed at near-infrared magnitudes, from {\it Spitzer}, {\it WHT} and {\it NTT}
data; these can be related to the total stellar mass with relatively
weak dependence on model assumptions.

We conclude the
following:
\begin{itemize}
\item The group population contains a clear red-sequence of galaxies
  down to the lowest stellar mass for which we are complete, $M\approx1\times
  10^{10}M_\odot$.  This red sequence is populated mostly by galaxies
  with elliptical or S0 morphology.  The most massive group galaxies,
  with $^{0.4}M_k<-23.5$ (corresponding to $M>1\times
  10^{11}M_\odot$), are predominantly found on the red sequence.
\item The fraction of galaxies in the red population depends on both
  near-infrared luminosity and environment. In all environments, the
  fraction of red galaxies decreases with decreasing luminosity, though
  the trend is relatively weak when near-infrared luminosities are
  used.  Nonetheless, at all luminosities the fraction of red galaxies in groups
  is significantly larger than in the field, by $\sim 20\pm 7$ per cent.  
\item In both the group and field samples, about 20 per cent of the optically-red galaxy
  population shows signs of dust-reddened star formation when longer
  wavelength photometry is considered.  Accounting for this,
  $54\pm6$ per cent of the group population brighter than
  $^{0.4}M_k=-22$ is evolving passively.  Only $35\pm 3$ per
  cent of the field population is comprised of such galaxies, and this is
  consistent with the number expected if they were {\it only}
  found within groups and clusters (which are proportionally
  represented in our field sample).
\item Considering fairly bright galaxies,  ${^{0.4}M_i}\leq-21.1$, we
  compare our group galaxies to a similar sample of cluster galaxies,
  and find that the fraction of red galaxies is intermediate between
  that of the field ($\sim 40$ per cent) and cluster ($\sim 80$ per
  cent) populations.  
\item Including comparably--selected groups at $z\sim 0.05$
  \citep{Weinmann_1+06,BB04}, we detect no significant redshift evolution in the red fraction of the
  groups since $z<0.55$, considering only the brightest galaxies for which we have a statistically complete
  sample, ${^{0.4}M_i}<-21.6$.   
\item The dynamics of the active and passive populations in the groups
  are not statistically distinguishable.  The velocity distributions, though statistically consistent
  with Gaussian distributions, show some flattening (and bimodality in
  the blue and dusty-star forming population) characteristic of
  infall.  
\item The optical colour distribution of these groups is reasonably
  well-matched by the predictions of simple galaxy formation models,
  although the models overpredict the relative number of red (inactive)
  galaxies in groups.  In particular, the \citet{Bower05} model predicts a
  red fraction of 75\%, considerably larger than the observed $56\pm 8$\%.
  The more recent model of \citet{Font+08} tries to address this by
  increasing the timescale required for satellite quenching.  While
  this model achieves a lower red fraction of 67\%, it
  distorts the overall colour distribution, filling in the colour-space
  between the red and blue peaks in a way that is not seen in the observations.
\end{itemize}

In summary, galaxy groups at $z=0.4$ appear to host galaxy populations
that are intermediate in their star-formation history distribution, between that of the
field and that of rich clusters at this redshift. This difference is
characterised by a reduced fraction of galaxies of a given stellar mass
undergoing current star formation activity, compared with the field.  The
colour distribution, and its dependence on stellar mass and
environment, is reasonably well reproduced by simple models that
assume satellite galaxies lose their hot corona of gas on relatively
short timescales.  However, it remains an unsolved challenge to
simultaneously match the red fraction and distinct bimodality of the
galaxy colour distribution in different environments.

\section{Acknowledgments}\label{sec-akn}
We thank the GALFORM team for making their model
predictions publicly available, and the CNOC2 team for allowing us to
use their unpublished redshifts.  We acknowledge useful
conversations with Andreea Font, Ian McCarthy, David Gilbank, Mike
Hudson, Frank van den Bosch and
Simone Weinmann which helped to shape this work.  We are also grateful
to the anonymous referee, who made several suggestions which improved
our analysis of these results.
This work is based on observations made with: the Spitzer Space
Telescope, operated by the Jet Propulsion Laboratory, California
Institute of Technology, under a contract with NASA; the NASA/ESA Hubble Space
Telescope, at the Space Telescope Science Institute, which is operated
by the Association of Universities for Research in Astronomy, Inc.,
under NASA contract NAS 5-26555; the Canada-France-Hawaii Telescope, operated by the
National Research Council (NRC) of Canada, the Institut National des
Science de l'Universe of the Centre National de la Recherche
Scientifique (CNRS) of France, and the University of Hawaii; the ESO
telescopes at the La Silla Observatories, under programme ids 076.A.0346 and
  077.A.0224; and the William Herschel Telescope, operated on the
  island of La Palma by t he Isaac Newton Group in the Spanish
  Observatorio del Roque de Los Muchachos of the Instituto de
  Astrofisica de Canarias.
This
research made use of tools provided by astrometry.net, and the facilities of the Canadian Astronomy Data 
Centre, which is operated by the National Research Council of Canada with 
the support of the Canadian Space Agency.  
MLB acknowledges financial support from an NSERC Discovery Grant.
\bibliography{ms}

\begin{thebibliography}{88}
\expandafter\ifx\csname natexlab\endcsname\relax\def\natexlab#1{#1}\fi

\bibitem[{{Baldry} {et~al.}(2006){Baldry}, {Balogh}, {Bower}, {Glazebrook},
  {Nichol}, {Bamford}, \& {Budavari}}]{BaldryV}
{Baldry}, I.~K., {Balogh}, M.~L., {Bower}, R.~G., {Glazebrook}, K., {Nichol},
  R.~C., {Bamford}, S.~P., \& {Budavari}, T. 2006, MNRAS, 373, 469

\bibitem[{{Baldry} {et~al.}(2004){Baldry}, {Glazebrook}, {Brinkmann},
  {Ivezi\'{c}}, {Lupton}, {Nichol}, \& {Szalay}}]{Baldry03}
{Baldry}, I.~K., {Glazebrook}, K., {Brinkmann}, J., {Ivezi\'{c}}, Z., {Lupton},
  R.~H., {Nichol}, R.~C., \& {Szalay}, A.~S. 2004, ApJ, 600, 681

\bibitem[{{Balogh} {et~al.}(2004{\natexlab{a}}){Balogh}, {Baldry}, {Nichol},
  {Miller}, {Bower}, \& {Glazebrook}}]{BB04}
{Balogh}, M.~L., {Baldry}, I.~K., {Nichol}, R.~C., {Miller}, C., {Bower},
  R.~G., \& {Glazebrook}, K. 2004{\natexlab{a}}, ApJL, 615, L101

\bibitem[{{Balogh} {et~al.}(1997){Balogh}, {Morris}, {Yee}, {Carlberg}, \&
  {Ellingson}}]{B+97}
{Balogh}, M.~L., {Morris}, S.~L., {Yee}, H.~K.~C., {Carlberg}, R.~G., \&
  {Ellingson}, E. 1997, ApJL, 488, L75+

\bibitem[{{Balogh} {et~al.}(1999){Balogh}, {Morris}, {Yee}, {Carlberg}, \&
  {Ellingson}}]{PSG}
---. 1999, ApJ, 527, 54

\bibitem[{{Balogh} {et~al.}(2000){Balogh}, {Navarro}, \& {Morris}}]{infall}
{Balogh}, M.~L., {Navarro}, J.~F., \& {Morris}, S.~L. 2000, ApJ, 540, 113

\bibitem[{{Balogh} {et~al.}(2004{\natexlab{b}})}]{2dfsdss}
{Balogh}, M.~L. {et~al.} 2004{\natexlab{b}}, MNRAS, 348, 1355

\bibitem[{{Balogh} {et~al.}(2007)}]{cnoc2_ir}
---. 2007, MNRAS, 374, 1169

\bibitem[{{Bamford} {et~al.}(2007){Bamford}, {Milvang-Jensen}, \&
  {Arag{\'o}n-Salamanca}}]{BMA}
{Bamford}, S.~P., {Milvang-Jensen}, B., \& {Arag{\'o}n-Salamanca}, A. 2007,
  MNRAS, 378, L6

\bibitem[{{Beers} {et~al.}(1990){Beers}, {Flynn}, \& {Gebhardt}}]{Beers}
{Beers}, T.~C., {Flynn}, K., \& {Gebhardt}, K. 1990, AJ, 100, 32

\bibitem[{{Bell} {et~al.}(2004){Bell}, {Wolf}, {Meisenheimer}, {Rix}, {Borch},
  {Dye}, {Kleinheinrich}, {Wisotzki}, \& {McIntosh}}]{BellGems2}
{Bell}, E.~F., {Wolf}, C., {Meisenheimer}, K., {Rix}, H., {Borch}, A., {Dye},
  S., {Kleinheinrich}, M., {Wisotzki}, L., \& {McIntosh}, D.~H. 2004, ApJ, 608,
  752

\bibitem[{{Bell} {et~al.}(2007{\natexlab{a}}){Bell}, {Zheng}, {Papovich},
  {Borch}, {Wolf}, \& {Meisenheimer}}]{Bell+07}
{Bell}, E.~F., {Zheng}, X.~Z., {Papovich}, C., {Borch}, A., {Wolf}, C., \&
  {Meisenheimer}, K. 2007{\natexlab{a}}, ApJ, 663, 834

\bibitem[{{Bell} {et~al.}(2007{\natexlab{b}})}]{Bell+07_short}
{Bell}, E.~F. {et~al.} 2007{\natexlab{b}}, ApJ, 663, 834

\bibitem[{{Berrier} {et~al.}(2009){Berrier}, {Stewart}, {Bullock}, {Purcell},
  {Barton}, \& {Wechsler}}]{Berr+08}
{Berrier}, J.~C., {Stewart}, K.~R., {Bullock}, J.~S., {Purcell}, C.~W.,
  {Barton}, E.~J., \& {Wechsler}, R.~H. 2009, ApJ, 690, 1292

\bibitem[{Bertin \& Arnouts(1996)}]{sextractor}
Bertin, E. \& Arnouts, S. 1996, A\&AS, 117, 393

\bibitem[{{Bildfell} {et~al.}(2008){Bildfell}, {Hoekstra}, {Babul}, \&
  {Mahdavi}}]{BHBM08}
{Bildfell}, C., {Hoekstra}, H., {Babul}, A., \& {Mahdavi}, A. 2008, MNRAS, 389,
  1637

\bibitem[{{Birnboim} \& {Dekel}(2003)}]{BD03}
{Birnboim}, Y. \& {Dekel}, A. 2003, MNRAS, 345, 349

\bibitem[{{Blanton} {et~al.}(2005){Blanton}, {Eisenstein}, {Hogg}, {Schlegel},
  \& {Brinkmann}}]{Blanton04}
{Blanton}, M.~R., {Eisenstein}, D., {Hogg}, D.~W., {Schlegel}, D.~J., \&
  {Brinkmann}, J. 2005, ApJ, 629, 143

\bibitem[{{Blanton} \& {Roweis}(2007)}]{kcorrect}
{Blanton}, M.~R. \& {Roweis}, S. 2007, AJ, 133, 734

\bibitem[{{Bower} {et~al.}(2006){Bower}, {Benson}, {Malbon}, {Helly}, {Frenk},
  {Baugh}, {Cole}, \& {Lacey}}]{Bower05}
{Bower}, R.~G., {Benson}, A.~J., {Malbon}, R., {Helly}, J.~C., {Frenk}, C.~S.,
  {Baugh}, C.~M., {Cole}, S., \& {Lacey}, C.~G. 2006, MNRAS, 370, 645

\bibitem[{{Bruzual} \& {Charlot}(2003)}]{BC03}
{Bruzual}, G. \& {Charlot}, S. 2003, MNRAS, 344, 1000

\bibitem[{{Carlberg} {et~al.}(2001{\natexlab{a}}){Carlberg}, {Yee}, {Morris},
  {Lin}, {Hall}, {Patton}, {Sawicki}, \& {Shepherd}}]{CNOC_groups2}
{Carlberg}, R.~G., {Yee}, H.~K.~C., {Morris}, S.~L., {Lin}, H., {Hall}, P.~B.,
  {Patton}, D.~R., {Sawicki}, M., \& {Shepherd}, C.~W. 2001{\natexlab{a}}, ApJ,
  563, 736

\bibitem[{{Carlberg} {et~al.}(2001{\natexlab{b}}){Carlberg}, {Yee}, {Morris},
  {Lin}, {Hall}, {Patton}, {Sawicki}, \& {Shepherd}}]{CNOC_groups}
---. 2001{\natexlab{b}}, ApJ, 552, 427

\bibitem[{{Cassata} {et~al.}(2008)}]{Cassata08}
{Cassata}, P. {et~al.} 2008, A\&A, 483, L39

\bibitem[{{Chabrier}(2003)}]{Chab}
{Chabrier}, G. 2003, PASP, 115, 763

\bibitem[{{Cole} {et~al.}(2000){Cole}, {Lacey}, {Baugh}, \& {Frenk}}]{Cole2000}
{Cole}, S., {Lacey}, C.~G., {Baugh}, C.~M., \& {Frenk}, C.~S. 2000, MNRAS, 319,
  168

\bibitem[{{Cooper} {et~al.}(2006)}]{DEEPII_envt}
{Cooper}, M.~C. {et~al.} 2006, MNRAS, 370, 198

\bibitem[{{Cooper} {et~al.}(2008)}]{DEEPII_envt_again}
---. 2008, MNRAS, 383, 1058

\bibitem[{{Cowie} \& {Barger}(2008)}]{CB08}
{Cowie}, L.~L. \& {Barger}, A.~J. 2008, ApJ, 686, 72

\bibitem[{{Crawford} {et~al.}(1999){Crawford}, {Allen}, {Ebeling}, {Edge}, \&
  {Fabian}}]{C+99}
{Crawford}, C.~S., {Allen}, S.~W., {Ebeling}, H., {Edge}, A.~C., \& {Fabian},
  A.~C. 1999, MNRAS, 306, 857

\bibitem[{{Croton} {et~al.}(2006)}]{Croton05}
{Croton}, D.~J. {et~al.} 2006, MNRAS, 365, 11

\bibitem[{{De Propris} {et~al.}(2003){De Propris}, {Stanford}, {Eisenhardt}, \&
  {Dickinson}}]{deP_BOK}
{De Propris}, R., {Stanford}, S.~A., {Eisenhardt}, P.~R., \& {Dickinson}, M.
  2003, ApJ, 598, 20

\bibitem[{{Dekel} \& {Birnboim}(2006)}]{DB06}
{Dekel}, A. \& {Birnboim}, Y. 2006, MNRAS, 368, 2

\bibitem[{{Edwards} {et~al.}(2007){Edwards}, {Hudson}, {Balogh}, \&
  {Smith}}]{Edwards+07}
{Edwards}, L.~O.~V., {Hudson}, M.~J., {Balogh}, M.~L., \& {Smith}, R.~J. 2007,
  MNRAS, 379, 100

\bibitem[{{Eke} {et~al.}(2004)}]{Eke-groups}
{Eke}, V.~R. {et~al.} 2004, MNRAS, 348, 866

\bibitem[{{Elbaz} {et~al.}(2007)}]{Elbaz+07}
{Elbaz}, D. {et~al.} 2007, A\&A, 468, 33

\bibitem[{{Ellingson} {et~al.}(2001){Ellingson}, {Lin}, {Yee}, \&
  {Carlberg}}]{Erica}
{Ellingson}, E., {Lin}, H., {Yee}, H.~K.~C., \& {Carlberg}, R.~G. 2001, ApJ,
  547, 609

\bibitem[{{Finn} {et~al.}(2008){Finn}, {Balogh}, {Zaritsky}, {Miller}, \&
  {Nichol}}]{Finn08}
{Finn}, R.~A., {Balogh}, M.~L., {Zaritsky}, D., {Miller}, C.~J., \& {Nichol},
  R.~C. 2008, ApJ, 679, 279

\bibitem[{{Font} {et~al.}(2008)}]{Font+08}
{Font}, A.~S. {et~al.} 2008, MNRAS, 389, 1619

\bibitem[{{Gallazzi} {et~al.}(2009)}]{Gallazzi+08}
{Gallazzi}, A. {et~al.} 2009, ApJ, 690, 1883

\bibitem[{{Gehrels}(1986)}]{Gehrels}
{Gehrels}, N. 1986, ApJ, 303, 336

\bibitem[{{Gerke} {et~al.}(2007)}]{Gerke06}
{Gerke}, B. {et~al.} 2007, MNRAS, 376, 1425

\bibitem[{{Gilbank} \& {Balogh}(2008)}]{GB08}
{Gilbank}, D.~G. \& {Balogh}, M.~L. 2008, MNRAS, 385, L116

\bibitem[{{Gilbank} {et~al.}(2003){Gilbank}, {Smail}, {Ivison}, \&
  {Packham}}]{Gilbank03}
{Gilbank}, D.~G., {Smail}, I., {Ivison}, R.~J., \& {Packham}, C. 2003, MNRAS,
  346, 1125

\bibitem[{{Gomez} {et~al.}(2003){Gomez}, {Nichol}, {et~al.}}]{Sloan_sfr_short}
{Gomez}, P.~L., {Nichol}, R.~C., {et~al.} 2003, ApJ, 584, 210

\bibitem[{{Gwyn}(2008)}]{Megapipe}
{Gwyn}, S.~D.~J. 2008, PASP, 120, 212

\bibitem[{{Haines} {et~al.}(2008){Haines}, {Gargiulo}, \& {Merluzzi}}]{HGM08}
{Haines}, C.~P., {Gargiulo}, A., \& {Merluzzi}, P. 2008, MNRAS, 385, 1201

\bibitem[{{Haines} {et~al.}(2006){Haines}, {La Barbera}, {Mercurio},
  {Merluzzi}, \& {Busarello}}]{Haines+06}
{Haines}, C.~P., {La Barbera}, F., {Mercurio}, A., {Merluzzi}, P., \&
  {Busarello}, G. 2006, ApJL, 647, L21

\bibitem[{{Hogg} {et~al.}(2008){Hogg}, {Blanton}, {Lang}, {Mierle}, \&
  {Roweis}}]{astrometrydotnet}
{Hogg}, D.~W., {Blanton}, M., {Lang}, D., {Mierle}, K., \& {Roweis}, S. 2008,
  in Astronomical Society of the Pacific Conference Series, Vol. 394,
  Astronomical Data Analysis Software and Systems XVII, ed. R.~W. {Argyle},
  P.~S. {Bunclark}, \& J.~R. {Lewis}, 27--+

\bibitem[{{Hsieh} {et~al.}(2005){Hsieh}, {Yee}, {Lin}, \&
  {Gladders}}]{Hsieh+05}
{Hsieh}, B.~C., {Yee}, H.~K.~C., {Lin}, H., \& {Gladders}, M.~D. 2005, ApJS,
  158, 161

\bibitem[{{Jarrett} {et~al.}(2000){Jarrett}, {Chester}, {Cutri}, {Schneider},
  {Skrutskie}, \& {Huchra}}]{2MASS}
{Jarrett}, T.~H., {Chester}, T., {Cutri}, R., {Schneider}, S., {Skrutskie}, M.,
  \& {Huchra}, J.~P. 2000, AJ, 119, 2498

\bibitem[{{Jeltema} {et~al.}(2007){Jeltema}, {Mulchaey}, {Lubin}, \&
  {Fassnacht}}]{JMLF}
{Jeltema}, T.~E., {Mulchaey}, J.~S., {Lubin}, L.~M., \& {Fassnacht}, C.~D.
  2007, ApJ, 658, 865

\bibitem[{{Juneau} {et~al.}(2005)}]{Juneau+04}
{Juneau}, S. {et~al.} 2005, ApJL, 619, L135

\bibitem[{{Kauffmann} {et~al.}(2003){Kauffmann}, {Heckman},
  {et~al.}}]{Kauffmann-SDSS1_short}
{Kauffmann}, G., {Heckman}, T.~M., {et~al.} 2003, MNRAS, 341, 54

\bibitem[{{Koopmann} \& {Kenney}(2004)}]{KK04}
{Koopmann}, R.~A. \& {Kenney}, J.~D.~P. 2004, ApJ, 613, 866

\bibitem[{{Lang} {et~al.}(2009){Lang}, {Hogg}, {Mierle}, {Blanton}, \&
  {Roweis}}]{lang09}
{Lang}, D., {Hogg}, D.~W., {Mierle}, K., {Blanton}, M., \& {Roweis}, S. 2009,
  in prep.

\bibitem[{{Lewis} {et~al.}(2002)}]{2dF_short}
{Lewis}, I.~J. {et~al.} 2002, MNRAS, 333, 279

\bibitem[{{Marchesini} {et~al.}(2008){Marchesini}, {van Dokkum}, {Forster
  Schreiber}, {Franx}, {Labbe'}, \& {Wuyts}}]{M+08}
{Marchesini}, D., {van Dokkum}, P.~G., {Forster Schreiber}, N.~M., {Franx}, M.,
  {Labbe'}, I., \& {Wuyts}, S. 2008, ArXiv e-prints

\bibitem[{{McCarthy} {et~al.}(2008){McCarthy}, {Frenk}, {Font}, {Lacey},
  {Bower}, {Mitchell}, {Balogh}, \& {Theuns}}]{Ian_rps}
{McCarthy}, I.~G., {Frenk}, C.~S., {Font}, A.~S., {Lacey}, C.~G., {Bower},
  R.~G., {Mitchell}, N.~L., {Balogh}, M.~L., \& {Theuns}, T. 2008, MNRAS, 383,
  593

\bibitem[{{McGee} {et~al.}(2009){McGee}, {Balogh}, {Bower}, {Font}, \&
  {McCarthy}}]{McGee-accretion}
{McGee}, S.~L., {Balogh}, M.~L., {Bower}, R.~G., {Font}, A., \& {McCarthy}, I.
  2009, MNRAS, submitted

\bibitem[{{McGee} {et~al.}(2008){McGee}, {Balogh}, {Henderson}, {Wilman},
  {Bower}, {Mulchaey}, \& {Oemler}}]{McGee}
{McGee}, S.~L., {Balogh}, M.~L., {Henderson}, R.~D.~E., {Wilman}, D.~J.,
  {Bower}, R.~G., {Mulchaey}, J.~S., \& {Oemler}, A.~J. 2008, MNRAS, 664

\bibitem[{{Moran} {et~al.}(2006){Moran}, {Ellis}, {Treu}, {Salim}, {Rich},
  {Smith}, \& {Kneib}}]{Moran+06}
{Moran}, S.~M., {Ellis}, R.~S., {Treu}, T., {Salim}, S., {Rich}, R.~M.,
  {Smith}, G.~P., \& {Kneib}, J.-P. 2006, ApJL, 641, L97

\bibitem[{{Muzzin} {et~al.}(2008){Muzzin}, {Wilson}, {Lacy}, {Yee}, \&
  {Stanford}}]{Muzzin+08}
{Muzzin}, A., {Wilson}, G., {Lacy}, M., {Yee}, H.~K.~C., \& {Stanford}, S.~A.
  2008, ApJ, 686, 966

\bibitem[{{Nakata} {et~al.}(2005){Nakata}, {Bower}, {Balogh}, \&
  {Wilman}}]{Nakata}
{Nakata}, F., {Bower}, R.~G., {Balogh}, M.~L., \& {Wilman}, D.~J. 2005, MNRAS,
  357, 679

\bibitem[{{O'Dea} {et~al.}(2008)}]{OD+08}
{O'Dea}, C.~P. {et~al.} 2008, ApJ, 681, 1035

\bibitem[{{Parker} {et~al.}(2005){Parker}, {Hudson}, {Carlberg}, \&
  {Hoekstra}}]{PHCH}
{Parker}, L.~C., {Hudson}, M.~J., {Carlberg}, R.~G., \& {Hoekstra}, H. 2005,
  ApJ, 634, 806

\bibitem[{{Pasquali} {et~al.}(2009){Pasquali}, {van den Bosch}, {Mo}, {Yang},
  \& {Somerville}}]{P+08}
{Pasquali}, A., {van den Bosch}, F.~C., {Mo}, H.~J., {Yang}, X., \&
  {Somerville}, R. 2009, MNRAS, 394, 38

\bibitem[{{Poggianti} {et~al.}(2006)}]{Pogg05}
{Poggianti}, B. {et~al.} 2006, ApJ, 642, 188

\bibitem[{{Poggianti} {et~al.}(2008)}]{Pogg+08}
{Poggianti}, B.~M. {et~al.} 2008, ApJ, 684, 888

\bibitem[{{Rafferty} {et~al.}(2008){Rafferty}, {McNamara}, \& {Nulsen}}]{RMN08}
{Rafferty}, D.~A., {McNamara}, B.~R., \& {Nulsen}, P.~E.~J. 2008, ApJ, 687, 899

\bibitem[{{Schlegel} {et~al.}(1998){Schlegel}, {Finkbeiner}, \& {Davis}}]{SFD}
{Schlegel}, D.~J., {Finkbeiner}, D.~P., \& {Davis}, M. 1998, ApJ, 500, 525

\bibitem[{{Sijacki} {et~al.}(2007){Sijacki}, {Springel}, {di Matteo}, \&
  {Hernquist}}]{Sijacki07}
{Sijacki}, D., {Springel}, V., {di Matteo}, T., \& {Hernquist}, L. 2007, MNRAS,
  380, 877

\bibitem[{{Somerville} {et~al.}(2008){Somerville}, {Hopkins}, {Cox},
  {Robertson}, \& {Hernquist}}]{Somerville08}
{Somerville}, R.~S., {Hopkins}, P.~F., {Cox}, T.~J., {Robertson}, B.~E., \&
  {Hernquist}, L. 2008, MNRAS, 1241

\bibitem[{{Strateva} {et~al.}(2001){Strateva}, {Ivezi{\' c}},
  {et~al.}}]{Strateva01_short}
{Strateva}, I., {Ivezi{\' c}}, {\v Z}., {et~al.} 2001, AJ, 122, 1861

\bibitem[{{Taylor} {et~al.}(2009)}]{Taylor+08}
{Taylor}, E.~N. {et~al.} 2009, ApJ, 694, 1171

\bibitem[{{van den Bosch} {et~al.}(2008)}]{vdB07_short}
{van den Bosch}, F.~C. {et~al.} 2008, MNRAS, 387, 79

\bibitem[{{Weinmann} {et~al.}(2006{\natexlab{a}}){Weinmann}, {van den Bosch},
  {Yang}, \& {Mo}}]{Weinmann_1+06}
{Weinmann}, S.~M., {van den Bosch}, F.~C., {Yang}, X., \& {Mo}, H.~J.
  2006{\natexlab{a}}, MNRAS, 366, 2

\bibitem[{{Weinmann} {et~al.}(2006{\natexlab{b}})}]{Weinmann+06_short}
{Weinmann}, S.~M. {et~al.} 2006{\natexlab{b}}, MNRAS, 372, 1161

\bibitem[{Whitaker(2005)}]{Whitaker}
Whitaker, R. 2005, PhD thesis, University of Durham

\bibitem[{{White} \& {Frenk}(1991)}]{WF91}
{White}, S.~D.~M. \& {Frenk}, C.~S. 1991, ApJ, 379, 52

\bibitem[{{Wilman} {et~al.}(2005{\natexlab{a}}){Wilman}, {Balogh}, {Bower},
  {Mulchaey}, {Oemler}, {Carlberg}, {Morris}, \& {Whitaker}}]{CNOC2_groupsI}
{Wilman}, D.~J., {Balogh}, M.~L., {Bower}, R.~G., {Mulchaey}, J.~S., {Oemler},
  A., {Carlberg}, R.~G., {Morris}, S.~L., \& {Whitaker}, R.~J.
  2005{\natexlab{a}}, MNRAS, 358, 71

\bibitem[{{Wilman} {et~al.}(2009){Wilman}, {Oemler}, {Mulchaey}, {McGee},
  {Balogh}, \& {Bower}}]{Wilman_morph}
{Wilman}, D.~J., {Oemler}, A., {Mulchaey}, J.~S., {McGee}, S.~L., {Balogh},
  M.~L., \& {Bower}, R.~G. 2009, ApJ, 692, 298

\bibitem[{{Wilman} {et~al.}(2005{\natexlab{b}})}]{CNOC2_groupsII}
{Wilman}, D.~J. {et~al.} 2005{\natexlab{b}}, MNRAS, 358, 88

\bibitem[{{Wilman} {et~al.}(2008)}]{cnoc2_irac}
---. 2008, ApJ, 680, 1009

\bibitem[{{Wolf} {et~al.}(2005){Wolf}, {Gray}, \& {Meisenheimer}}]{C17-dust}
{Wolf}, C., {Gray}, M.~E., \& {Meisenheimer}, K. 2005, A\&A, 443, 435

\bibitem[{{Wolf} {et~al.}(2009)}]{STAGES-dust}
{Wolf}, C. {et~al.} 2009, MNRAS, 128

\bibitem[{{Yang} {et~al.}(2007){Yang}, {Mo}, {van den Bosch}, {Pasquali}, {Li},
  \& {Barden}}]{Y07}
{Yang}, X., {Mo}, H.~J., {van den Bosch}, F.~C., {Pasquali}, A., {Li}, C., \&
  {Barden}, M. 2007, ApJ, 671, 153

\bibitem[{{Yee} {et~al.}(2000){Yee}, {Morris}, {Lin}, {Carlberg},
  {et~al.}}]{CNOC2-I_short}
{Yee}, H.~K.~C., {Morris}, S.~L., {Lin}, H., {Carlberg}, R.~G., {et~al.} 2000,
  ApJS, 129, 475

\end{thebibliography}
\end{document}